\documentclass[12pt,a4paper,dvips]{article}
\usepackage{amsmath,graphicx}

\newcommand{\pr}{\rightarrow}

\newcommand{\ba}{\begin{array}}
\newcommand{\ea}{\end{array}}

\newcommand{\vart}{\vartheta}
\newcommand{\varp}{\varphi}
\newcommand{\eps}{\varepsilon}

\newcommand{\il}{\int\limits}
\newenvironment{inspring}[1]%
{\begin{list}{}{\setlength{\rightmargin}{0cm}
                \setlength{\listparindent}{0cm}
                \settowidth{\labelwidth}{\mbox{#1}}
                \setlength{\leftmargin}{1.1\labelwidth}
                \setlength{\labelsep}{.1\labelwidth}}}%
{\end{list}}
\newcommand{\ITEM}[1]{\item[#1\hfill]}
\newcommand{\bi}[1]{\begin{inspring}{#1}}
\newcommand{\ei}{\end{inspring}}

\newcommand{\dil}{\displaystyle \int\limits}

\newcommand{\bfr}{{\bf r}}

\newcommand{\beq}{\begin{equation}}
\newcommand{\eq}{\end{equation}}
\catcode`\@=11

\font\tenmsa=msam10 \font\sevenmsa=msam7 \font\fivemsa=msam5
\font\tenmsb=msbm10 \font\sevenmsb=msbm7 \font\fivemsb=msbm5
\newfam\msafam
\newfam\msbfam
\textfont\msafam=\tenmsa  \scriptfont\msafam=\sevenmsa
  \scriptscriptfont\msafam=\fivemsa
\textfont\msbfam=\tenmsb  \scriptfont\msbfam=\sevenmsb
  \scriptscriptfont\msbfam=\fivemsb

\def\Bbb{\ifmmode\let\next\Bbb@\else
 \def\next{\errmessage{Use \string\Bbb\space only in math mode}}\fi\next}
\def\Bbb@#1{{\Bbb@@{#1}}}
\def\Bbb@@#1{\fam\msbfam#1}
\newcommand{\dR}{{\Bbb R}}

\title{A generalization of the Zernike circle polynomials for forward and inverse problems in diffraction theory}
\author{\mbox{} \\ A.J.E.M.\ Janssen \\
Eindhoven University of Technology, \\
Department EE and EURANDOM, LG 1.39, \\
P.O.\ Box 513, 5600 MB Eindhoven, \\
The Netherlands \\
e-mail: a.j.e.m.janssen@tue.nl}
\date{}

\begin{document}
\maketitle
\mbox{} \\ \\ \\ \\
\noindent
{\bf Abstract.} \\
A generalization of the Zernike circle polynomials for expansion of functions vanishing outside the unit disk is given. These generalized Zernike functions have the form $Z_n^{m,\alpha}(\rho,\vart)=R_n^{m,\alpha}(\rho)\,\exp(im\vart)$, $0\leq\rho<1$, $0\leq\vart<2\pi$, and vanish for $\rho>1$, where $n$ and $m$ are integers such that $n-|m|$ is non-negative and even. The radial parts are $O((1-\rho^2)^{\alpha})$ as $\rho\uparrow1$ in which $\alpha$ is a real parameter $>\:{-}1$. The $Z_n^{m,\alpha}$ are orthogonal on the unit disk with respect to the weight function $(1-\rho^2)^{-\alpha}$, $0\leq\rho<1$. The Fourier transform of $Z_n^{m,\alpha}$ can be expressed explicitly in terms of (generalized) Jinc functions $J_{n+\alpha+1}(2\pi r)/(2\pi r)^{\alpha+1}$ and exhibits a decay behaviour $r^{-\alpha-3/2}$ as $r\pr\infty$. This explicit result for the Fourier transform generalizes the basic identity in the classical Nijboer-Zernike theory, case $\alpha=0$, of optical diffraction. The generalized Zernike functions were considered by Tango in 1977 and a version of the result on the Fourier transform is presented, however, in somewhat different form and without a detailed proof. The new functions accommodate the solution of various forward and inverse problems in diffraction theory and related fields in which functions vanishing outside the unit disk with prescribed behaviour at the edge of the disk are involved. Furthermore, the explicit result for the Fourier transform of the $Z_n^{m,\alpha}$ allows formulation and solution of design problems, involving functions vanishing outside the unit disk whose Fourier transforms and their decay are prescribed, on the level of expansion coefficients with respect to the $Z_n^{m,\alpha}$. Cormack's result for the Radon transform of the classical circle polynomials admits an explicit generalization in which the Gegenbauer polynomials $C_n^{\alpha+1}$ appear. This result can be used to express the radial parts $R_n^{m,\alpha}(\rho)$ as Fourier coefficients of $C_n^{\alpha+1}(\rho\cos\vart)$ and to devise a computation scheme of the DCT-type for fast and reliable computation of the $R_n^{m,\alpha}(\rho)$.

In recent years, several (semi-) analytic computation schemes for forward and inverse problems in optical (ENZ) and acoustic (ANZ) diffraction theory have been developed for the classical case $\alpha=0$. Many of these schemes admit a generalization to the cases that $\alpha\neq0$. For instances that this is not the case, a connection formula expressing the generalized Zernike functions as a linear combination of the classical ones can be used. Various acoustic quantities that can be expressed via King's integral (involving the Hankel transform of radially symmetric functions) can be expressed for the general case that $\alpha\neq0$ in a form that generalizes the form that holds for the case $\alpha=0$. Furthermore, an inverse problem of estimating a velocity profile on a circular, baffled piston can be solved in terms of expansion coefficients from near-field data using Weyl's representation result for spherical waves. Finally, the new functions, case $\alpha={\pm}1/2$, are compared with certain non-orthogonal trial functions as used by Streng and by Mellow and K\"arkk\"ainen with respect to their ability in solving certain design problems in acoustic radiation with boundary conditions on a disk.
\newpage
\noindent
\section{Introduction} \label{sec1}
\mbox{} \\[-9mm]

In optical and acoustic radiation with boundary conditions given on a disk, the choice of basis functions to represent the boundary conditions is of great importance. Here one can consider the following requirements:
\bi{A.0}
\ITEM{A.} the basis functions should be appropriate given the physical background,
\ITEM{B.} the basis functions should be effective and accurate in their ability of representing functions vanishing outside the disk and arising in the particular physical context,
\ITEM{C.} the basis functions should have convenient analytic results for transformations that arise in a natural way in the given physical context.
\ei
For the case of optical diffraction with a circular pupil having both phase and amplitude non-uniformities, Zernike \cite{ref1} introduced in 1934 his circle polynomials, denoted here as $Z_n^m(\rho,\vart)$ with radial variable $\rho\geq0$ and angular variable $\vart$, vanishing for $\rho>1$, which find nowadays wide-spread application in fields like optical engineering and lithography \cite{ref2}--\cite{ref7}, astronomy \cite{ref8}--\cite{ref10} and ophthalmology \cite{ref11}--\cite{ref13}. The Zernike circle polynomials are given for integer $n$ and $m$ such that $n-|m|$ is even and non-negative as
\begin{eqnarray} \label{e1}
Z_n^m(\rho,\vart) & = & R_n^{|m|}(\rho)\,e^{im\vart}~,~~~~~~0\leq\rho<1\,,~~0\leq\vart<2\pi~, \nonumber \\[2mm]
& = & 0\hspace*{2cm},~~~~~~\rho>1~,
\end{eqnarray}
where the radial polynomials $R_n^{|m|}$ are given by
\beq \label{e2}
R_n^{|m|}(\rho)=\rho^{|m|}\,P_{\frac{n-|m|}{2}}^{(0,|m|)}\,(2\rho^2-1)~,
\eq
with
$P_k^{(\alpha,\beta)}$ the general Jacobi polynomial as given in \cite{ref14}, Ch.~22, \cite{ref15}, Ch.~4 and \cite{ref16}, Ch.~5, \S4. The circle polynomials were investigated by Bhatia and Wolf \cite{ref17} with respect to their appropriateness for use in optical diffraction theory, see issue A above, and were shown to arise more or less uniquely as orthogonal functions with polynomial radial dependence satisfying form invariance under rotations of the unit disk.

The orthogonality condition,
\beq \label{e3}
\il_0^1\il_0^{2\pi}\,Z_{n_1}^{m_1}(\rho,\vart)(Z_{n_2}^{m_2}(\rho,\vart))^{\ast}\,\rho\,d\rho\,d\vart=\frac{\pi} {n+1}\,\delta_{m_1m_2}\,\delta_{n_1n_2}~,
\eq
where $n$ is either one of $n_1$ and $n_2$ at the right-hand side, together with the completeness, see \cite{ref18}, App.~VII, end of Sec.~I, guarantees effective and accurate representation of square integrable functions on the unit disk in terms of their expansion coefficients with respect to the $Z_n^m$, see issue B above. The Zernike circle polynomials, notably those of azimuthal order $m=0$, were considered recently by Aarts and Janssen \cite{ref19}--\cite{ref22} for use in solving a variety of forward and inverse problems in acoustic radiation. This raised in the acoustic community \cite{ref23} the question how the circle polynomials with $m=0$ compare to other sets of non-polynomial, radially symmetric, orthogonal functions on the disk. It was shown in \cite{ref24} that the expansion coefficients, when using circle polynomials, properly reflect smoothness of the functions to be expanded in terms of decay and that they compare favourably in this respect with the expansion coefficients that occur when using orthogonal Bessel series expansions; the latter expansions are sometimes used both in the acoustic and the optical domain, see \cite{ref25}--\cite{refY}.

The present paper focuses on analytic properties, see issue C above, of basis functions, and in Sec.~\ref{sec2} we present a number of such properties for the set of Zernike circle polynomials. In Sec.~\ref{sec3} a generalization of the set of Zernike circle polynomials is introduced, viz.\ the set of functions
\begin{eqnarray} \label{e4}
Z_n^{m,\alpha}(\rho,\vart) & = & (1-\rho^2)^{\alpha}\,\rho^{|m|}\,P_{\frac{n-|m|}{2}}^{(\alpha,|m|)} (2\rho^2-1)\,e^{im\vart}~,~~~~~~0\leq\rho<1~, \nonumber \\[2mm]
& = & 0\hspace*{6.1cm},~~~~~~\rho>1~,
\end{eqnarray}
with $\rho$, $\vart$ and $n$, $m$ as before, see (\ref{e1}), and parameter $\alpha>{-}1$. These new Zernike-type functions arise sometimes more naturally in certain physical problems and can be more convenient when solving inverse problems in diffraction theory since the decay in the Fourier domain can be controlled by choosing the parameter $\alpha$ appropriately. While issue A above should guide the choice of $\alpha$, the matter of orthogonality and completeness is settled as in the classical case $\alpha=0$. A more involved question is what becomes of the analytic properties, noted for the classical circle polynomials in Sec.~\ref{sec2}, when $\alpha\neq0$. A major part of the present paper is concerned with answering this question, and this yields generalization of many of the results holding for the case $\alpha=0$ and about which more specifics will be given at the end of Sec.~\ref{sec3} where the basic properties of the generalized Zernike functions are presented.

\section{Analytic results for the classical circle polynomials} \label{sec2}
\mbox{} \\[-9mm]

The Zernike circle polynomials were used by Nijboer in his 1942 thesis \cite{ref27} for the computation of the point-spread functions in near best-focus planes pertaining to circular optical systems of low-to-medium numerical aperture. In that case, starting from a non-uniform pupil function
\beq \label{e5}
P(\rho,\vart)=A(\rho,\vart)\,e^{i\Phi(\rho,\vart)}~,~~~~~~0\leq\rho<1\,,~~0\leq\vart<2\pi~,
\eq
the point-spread function $U(r,\varp\,;\,f)$ at defocus plane $f$ with polar coordinates $x+iy=r\,e^{i\varp}$ is given in accordance with well established practices in Fourier optics as
\beq \label{e6}
U(r,\varp\,;\,f)=\il_0^1\il_0^{2\pi}\,e^{if\rho^2}\,e^{2\pi i\rho r\cos(\vart-\varp)}\,P(\rho,\vart)\,\rho \,d\rho\,d\vart~.
\eq
It was discovered by Zernike \cite{ref1} that the Fourier transform of the circle polynomials,
\beq \label{e7}
({\cal F}\,Z_n^m)(r,\varp)=\il_0^1\il_0^{2\pi}\,e^{2\pi i\rho r\cos(\vart-\varp)}\,Z_n^m(\rho,\vart)\,\rho\,d\rho\,d\vart
\eq
has the closed-form result
\beq \label{e8}
({\cal F}\,Z_n^m)(r,\varp)=2\pi\,i^n\,\frac{J_{n+1}(2\pi r)}{2\pi r}\,e^{im\varp}~,~~~~~~r\geq0\,,~~0\leq\varp<2\pi~,
\eq
where $J_{n+1}$ is the Bessel function of the first kind and order $n+1$. Thus, one expands the pupil function $P$ as
\beq \label{e9}
P(\rho,\vart)=\sum_{n,m}\,\beta_n^m\,Z_n^m(\rho,\vart)~,~~~~~~0\leq\rho<1\,,~~0\leq\vart<2\pi~,
\eq
and uses the result in (\ref{e8}) to compute the point-spread function $U$ in (\ref{e6}) in terms of the expansion coefficients $\beta_n^m$. In the case of best focus, $f=0$, this gives a computation result for the point-spread function immediately. For small values of $f$, say $|f|\leq1$, one expands the focal factor,
\beq \label{e10}
e^{if\rho^2}=1+if\rho^2-\tfrac12\,f^2\rho^4-...
\eq
and spends some additional effort to write functions $\rho^{2k}\,Z_n^m(\rho,\vart)$ as a $k$-terms linear combination of circle polynomials with upper index $m$ to which (\ref{e7}) applies. Also see \cite{ref18}, Ch.~9, Secs.~2--4. These are the main features of the classical Nijboer-Zernike theory for computation of optical point-spread functions in the presence of aberrations.

In \cite{ref28} Janssen has computed the point-spread function $U_n^m(r,\varp\,;\,f)$ pertaining to a single term $Z_n^m$ in the form
\beq \label{e11}
U_n^m(r,\varp\,;\,f)=2\pi\,i^{|m|}\,V_n^{|m|}(r,f)\,e^{im\varp}~,
\eq
where
\begin{eqnarray} \label{e12}
V_n^{|m|}(r,f) & = & \il_0^1\,e^{if\rho^2}\,R_n^{|m|}(\rho)\,J_{|m|}(2\pi\rho r)\,\rho\,d\rho~= \nonumber \\[3.5mm]
& = & e^{if}\,\sum_{l=0}^{\infty}\,\Bigl(\frac{{-}if}{\pi r}\Bigr)^l\,\sum_{j=0}^p\, u_{lj}\, \frac{J_{|m|+l+2j+1}(2\pi r)}{2\pi r}
\end{eqnarray}
with explicitly given $u_{lj}$ ($p=\frac12\,(n-|m|)$). This result has led to what is called the extended Nijboer-Zernike (ENZ) theory of forward and inverse computation for optical aberrations. The theory in its present form allows computation of point-spread functions for general $f$ and for high-NA optical systems, including polarization and birefringence, as well as for multi-layer systems. See \cite{ref29}--\cite{ref33} and \cite{ref34} for an overview. Furthermore, it provides a framework for estimating pupil functions $P$, in terms of expansion coefficients $\beta$, from measured data $|U|^2$ of the intensity point-spread function in the focal region. The latter inverse problem has the basic assumption that $P$ deviates only mildly from being constant so that the term with $m=n=0$ in (\ref{e9}) dominates the totality of all other terms. The theoretical intensity point-spread function can then, with modest error, be linearized around the leading term $|\beta_0^0\,U_0^0|^2$, and this leads, via a matching procedure with the measured data in the focal region, to a first estimate of the unknown coefficients. This procedure is made iterative by incorporating the totality of all deleted small cross-terms (involving the $\beta_n^m$, $(m,n)\neq(0,0)$, quadratically) in the matching procedure using the estimate of the $\beta$'s from the previous steps. In practice one finds that pupil functions $P$ deviating from being constant by as much as 2.5 times the diffraction limit can be retrieved. See \cite{ref31}, \cite{ref32}, \cite{ref34}, \cite{ref35}, \cite{ref36} for more details.

The result in (\ref{e8}) is one evidence of computational appropriateness of the circle polynomials for forward and inverse problems, but there are others. Many of these are based on the basic NZ-result in (\ref{e8}). In \cite{ref37}--\cite{ref38}, Cormack used this result to calculate the Radon transform ${\cal R}_n^m$,
\beq \label{e13}
{\cal R}_n^m(\tau,\psi)=\il_{l(\tau,\psi)}\,Z_n^m(\nu,\mu)\,dl
\eq
of $Z_n^m$, with integration along the line $\nu\cos\psi+\mu\sin\psi=\tau$ in the plane with $\tau\geq0$ and $\psi\in[0,2\pi)$ and where $(\nu,\mu)=(\rho\cos\vart,\rho\sin\vart)$. The result is that
\beq \label{e14}
{\cal R}_n^m(\tau,\psi)=\frac{2}{n+1}\,(1-\tau^2)^{1/2}\,U_n(\tau)\,e^{im\psi}~,~~~~~0\leq\tau\leq1\,,~~0\leq\psi<2\pi\,,
\eq
where $U_n$ is the Chebyshev polynomial of the second kind and degree $n$, \cite{ref14}, Ch.~22. On this explicit form of the Radon transform of $Z_n^m$, Cormack based a method for estimating a function on the disk from its Radon transform by estimating its Zernike expansion coefficients through matching. (In 1979, Cormack was awarded the Nobel prize in medicine, together with Hounsfield, for their work in computerized tomography.) Cormack's result was used by Dirksen and Janssen \cite{ref39} to find the integral representation
\beq \label{e15}
R_n^m(\rho)=\frac{1}{2\pi}\,\il_0^{2\pi}\,U_n(\rho\cos\vart)\cos m\vart\,d\vart
\eq
(integer $m,n\geq0$) that displays, for any $\rho\geq0$, the value of the radial part at $\rho$ in the form of the Fourier coefficient of a trigonometric polynomial. This formula (\ref{e15}) can be discretized, error free when more that $m+n$ equidistant points in $[0,2\pi]$ are used, and this yields a scheme of the DCT-type for computation of $R_n^m(\rho)$.

A further consequence of the basic NZ-result (\ref{e8}) is the theory of shifted-and-scaled circle polynomials developed in \cite{ref40}. For given $a\geq0$, $b\geq0$ with $a+b\leq1$, there are developed explicit expressions for the coefficients $K_{nn'}^{mm'}(a,b)$ in the expansion
\beq \label{e16}
Z_n^m((a+b\rho'\cos\vart',b\rho'\sin\vart'))=
\sum_{n',m'}\,K_{nn'}^{mm'}(a,b)\,Z_{n'}^{m'}(\rho'\cos\vart,\rho'\sin\vart')~.
\eq
This generalizes the result in \cite{ref41},
\beq \label{e17}
R_n^m(\eps\rho)=\sum_{n'=m(2)n}\,(R_n^{n'}(\eps)-R_n^{n'+2}(\eps))\,R_{n'}^m(\rho)~,
\eq
on the Zernike expansion of scaled circle polynomials ($m\geq 0$).

The basic NZ-result is also useful for the computation of various acoustic quantities that arise from a circular piston in a planar baffle. The complex amplitude $p(\bfr,k)$ of the sound pressure at the field point $\bfr=(x,y,z)$, $z\geq0$, in front of the baffle plane $z=0$ due to a harmonic excitation $\exp(i\omega t)$ with wave number $k=\omega/c$, with $c$ the speed of sound, is given by Rayleigh's integral and King's integral as
\begin{eqnarray} \label{e18}
p(\bfr,k) & = & \frac{i\rho_0 ck}{2\pi}\,\il_S\,v(\sigma)\,\frac{e^{-ikr'}}{r'}\,dS~= \nonumber \\[3.5mm]
& = & i\rho_0ck\,\il_0^{\infty}\,\frac{e^{-z(u^2-k^2)^{1/2}}}{(u^2-k^2)^{1/2}}\,J_0(wu)\,V(u)\,u\,du~.
\end{eqnarray}
Here $\rho_0$ is the density of the medium, $v(\sigma)$ is a non-uniform velocity profile assumed to depend on the radial variable $\sigma=(\nu^2+\mu^2)^{1/2}$ on the piston surface $S$ with center ${\bf 0}$ and radius $a$, and $r'$ is the distance from the field point $\bfr$ to the point $(\nu,\mu,0)$ on $S$. Furthermore, $w=(x^2+y^2)^{1/2}$ is the distance from the field point $\bfr$ to the $z$-axis, the root $(u^2-k^2)^{1/2}$ has the value $i\,\sqrt{k^2-u^2}$ and $\sqrt{u^2-k^2}$ for $0\leq u\leq k$ and $u\geq k$, respectively with $\sqrt{~~}$ non-negative in both cases, and $V(u)$ is the Hankel transform,
\beq \label{e19}
V(u)=\il_0^a\,J_0(u\sigma)\,v(\sigma)\,\sigma\,d\sigma~,~~~~~~u\geq0~,
\eq
of order 0 of $v(\sigma)$.

In \cite{ref19}, the on-axis pressure $p_{2l}(\bfr=(0,0,z),k)$ due to $v(\sigma)=R_{2l}^0(\sigma/a)$, $0\leq\sigma\leq a$, was shown from Rayleigh's integral and a special result on spherical Bessel functions to be given as
\beq \label{e20}
p_{2l}(\bfr=(0,0,z),k)=\tfrac12\,\rho_0\,c(ka)^2({-}1)^l\,j_l(kr_-)\,h_l^{(2)}(kr_+)~,
\eq
where $j_l$ and $h_l^{(2)}$ are spherical Bessel functions, see \cite{ref14}, Ch.~10 and in particular 10.1.45--46, and $r_{\pm}=\frac12\,[(r^2+a^2)^{1/2}\pm r]$. This result was used in \cite{ref19} for estimating a velocity profile from on-axis pressure data on the level of expansion coefficients with respect to radially symmetric circle polynomials.

In \cite{ref20}, the King integral for the pressure is employed to express the pressure $p((1,0,0),k)$ at the edge, the reaction force $\il_S\,p\,dS$ and the total radiated power $\il_S\,p(0)\,v^{\ast}(\sigma)\,dS$ in integral form. Expanding $v(\sigma)$ into radially symmetric circle polynomials and using the basic NZ-result for an explicit expression  of the Hankel transform $V(u)$ in (\ref{e19}), this gives rise to integrals
\beq \label{e21}
\il_0^{\infty}\,\frac{J_m(au)\,J_{n+1}(au)}{(u^2-k^2)^{1/2}}\,du~,~~~~~~\il_0^{\infty}\, \frac{J_{m+1}(au)\,J_{n+1}(au)}{(u^2-k^2)^{1/2}\,u}\,du
\eq
with integer $m,n\geq0$ of same parity. These integrals have been evaluated as a power series in $ka$ in \cite{ref20}.

In \cite{ref21}, the problem of sound radiation from a flexible spherical cap on a rigid sphere is considered. The scaling theory of Zernike circle polynomials, appropriately warped so as to account for the spherical geometry of the problem, is used to bring the standard solution of the Helmholtz equation with axially symmetric boundary data in a semi-analytic form per warped Zernike term. This gives rise to a coefficient-based solution of the inverse problem of estimating an axially symmetric velocity profile on the cap from measured pressure data in the space around the sphere.

Finally, returning to baffled-piston radiation, in \cite{ref22} the impulse response $h(\bfr,t)$, $t\geq0$, at a field point $\bfr=(x,y,z)$ in front of the baffle is considered. This $h(\bfr,t)$ is obtained as a Fourier inversion integral with respect to wave number $k$ of the velocity potential $\varp(\bfr,k)=(i\rho_0ck)^{-1}\,p(\bfr,k)$, which can be evaluated, via Rayleigh's integral, as an integral of the velocity profile $v$ over all points in the baffle plane at a common distance $(c^2t^2-z^2)^{1/2}$ from the field point. For the case that $v(\sigma)=R_{2l}^0(\sigma/a)$, $0\leq\sigma\leq a$, the latter integral can be evaluated explicitly using the addition theorem for Legendre polynomials. This explicit result can be used to compute impulse responses for general $v$ by expanding such a $v$ into radially symmetric circle functions. And, of course, an inverse problem, with a coefficient-based solution, can again be formulated.

\section{Generalized Zernike circle functions} \label{sec3}
\mbox{} \\[-9mm]

The Zernike circle polynomials, see (\ref{e1}), all have modulus 1 at the edge $\rho=1$ of the unit disk while in both Optics and Acoustics it frequently occurs that the non-uniformity behaves differently towards the edge of the disk.

In optical design, it is often desirable to have a pupil function $P$ whose point-spread function $U$, see (\ref{e6}), decays relatively fast outside the focal region where the specifications of the designer are to be met. The Zernike circle polynomials are discontinuous at the edge $\rho=1$, and the corresponding point-spread functions $U_n^m(r,\varp\,;\,f)$ have poor decay, like $r^{-3/2}$ as $r\pr\infty$.

In the ENZ point-spread functions, see Sec.~\ref{sec2}, the key assumption is that the pupil function's deviation from being constant is not large. When $|P|$ gets small at the edge, which happens, for instance, when the source is a pinhole with a positive diameter while the objective lens has a large NA, this basic assumption is not met. In such a case, aberration retrieval with the ENZ method may become cumbersome, even in its iterative version.

In the theory of acoustic radiation from a baffled, planar piston with radially symmetric boundary conditions, Streng \cite{ref42} follows the approach of Bouwkamp \cite{ref43} in solving the Helmholtz equation for this case and postulates normalized pressure functions on the disk of the form
\beq \label{e22}
\sum_{n=0}^{\infty}\,a_n(1-\rho^2)^{n+1/2}~,~~~~~~0\leq\rho<1~.
\eq
In the light of the later developments of this paper, it is interesting to note here that Bouwkamp himself prefers expansions that involve the functions
\beq \label{e23}
(1-\rho^2)^{1/2}\,P_l^{(1/2,0)}(2\rho^2-1)=({-}1)^l\,P_{2l+1}((1-\rho^2)^{1/2})~,~~~~~~0\leq\rho<1~.
\eq
Similarly, Mellow \cite{ref44} has velocity profiles on the disk of the form (normalized to the unit disk)
\beq \label{e24}
v(\rho)=\sum_{n=0}^{\infty}\,b_n(1-\rho^2)^{n-1/2}~,~~~~~~0\leq\rho<1~.
\eq
In \cite{ref44} and \cite{ref45}--\cite{ref46}, the postulates (\ref{e22}), (\ref{e24}) are used to solve design problems in acoustic radiation from a membrane in a circular disk in terms of the expansion coefficients $a_n$, $b_n$. In principle, these design problems could also be solved when radially symmetric circle polynomials instead of $(1-\rho^2)^{n\pm1/2}$ were used as trial functions, but this would become cumbersome, the functions $(1-\rho^2)^{\pm1/2}$ themselves already having a poorly convergent expansion with respect to the $R_{2l}^0(\rho)$.

In the three instances just discussed, it would be quite helpful when the system of Zernike circle polynomials would be replaced by systems whose members exhibit an appropriate behaviour at the edge of the disk. These new systems should satisfy the requirements A, B and C mentioned in the beginning of Sec.~\ref{sec1}. In this paper, the set of functions
\begin{eqnarray} \label{e25}
Z_n^{m,\alpha}(\rho,\vart) & \!\!= & \!\!(1-\rho^2)^{\alpha}\,\rho^{|m|}\,P_{\frac{n-|m|}{2}}^{(\alpha,|m|)}(2\rho^2-1)\, e^{im\vart}~,~~~0\leq\rho<1\,,~0\leq\vart<2\pi \nonumber \\[3mm]
& \!\!= & \!\!0\hspace*{6.1cm},~~~\rho>1~,
\end{eqnarray}
is considered for this purpose. In (\ref{e25}), the parameter $\alpha>{-}1$, and $n$ and $m$ are integers such that $n-|m|$ is even and non-negative, and the $P_k^{(\alpha,\beta)}$ are Jacobi polynomials as before. Observe that the radial parts
\beq \label{e26}
R_n^{|m|,\alpha}(\rho)=(1-\rho^2)^{\alpha}\, \rho^{|m|}\, P_{\frac{n-|m|}{2}}^{(\alpha,|m|)}(2\rho^2-1)~,~~~~~~0\leq\rho<1~,
\eq
depend non-polynomially on $\rho$, unless $\alpha=0,1,...\,$. The radial parts occur essentially in Tango \cite{ref47}, Sec.~1, but there the factor $(1-\rho^2)^{\alpha}$ is replaced by $(1-\rho^2)^{\alpha/2}$ with somewhat different restrictions on $\alpha$ and $|m|$. There is also a relation with the disk polynomials as they occur, for instance, in the work of Koornwinder \cite{ref48}:
\beq \label{e27}
Z_n^{m,\alpha}(\rho,\vart)=(1-\rho^2)^{\alpha}\,D_{\frac{n+m}{2}\,,\,\frac{n-m}{2}}^{\alpha}(\rho\,e^{i\vart})~,
\eq
with
\beq \label{28}
D_{k,l}^{\alpha}(\rho\,e^{i\vart})=\rho^{|k-l|}\,P_{\min(k,l)}^{(\alpha,|k-l|)}(2\rho^2-1)\,e^{i(k-l)\vart}
\eq
for $k,l=0,1,...\,$. Thus, the disk polynomials omit the factor $(1-\rho^2)^{\alpha}$ altogether.

Further definitions used in this paper are
\beq \label{e29}
p=\frac{n-|m|}{2}~,~~~~~~q=\frac{n+|m|}{2}
\eq
for integers $n$ and $m$ such that $n-|m|$ is even and non-negative, and the generalized Pochhammer symbol
\beq \label{e30}
(x)_y=\frac{\Gamma(x+y)}{\Gamma(x)}~.
\eq

By orthogonality of the $P_k^{(\alpha,\beta)}(x)$ with respect to the weight function $(1-x)^{\alpha}(1+x)^{\beta}$ on $[{-}1,1]$, there holds
\begin{eqnarray} \label{e31}
& \mbox{} & \frac{1}{\pi}\,\il_0^1\il_0^{2\pi}\,(1-\rho^2)^{-\alpha}\,Z_{n_1}^{m_1,\alpha}(\rho,\vart) (Z_{n_2}^{m_2,\alpha}(\rho,\vart))^{\ast}\,\rho\,d\rho\,d\vart~= \nonumber \\[3.5mm]
& & =~\frac{(p+1)_{\alpha}}{(p+|m|+1)_{\alpha}}~\frac{\delta_{m_1m_2}\,\delta_{n_1n_2}}{n_1+\alpha+1}~.
\end{eqnarray}

We note that for any integer $N=0,1,...$
\beq \label{e32}
\rho^{|m|}\,P_{\frac{n-|m|}{2}}^{(\alpha,|m|)}(2\rho^2-1)\,e^{im\vart}
\eq
with $n=0,1,...,N$ and $m={-}n,{-}n+2,...,n$ are $\frac12\,(N+1)(N+2)$ linearly independent functions of the form $\sum_{i,j}\,a_{ij}\nu^i\mu^j$ ($\nu=\rho\cos\vart$, $\mu=\rho\sin\vart$) where the summation is over integer $i,j\geq0$ with $i+j\leq N$. Therefore, by Weierstrass theorem the functions in (\ref{e32}) are complete.

This paper focuses on establishing versions for the generalized Zernike functions of the analytic results that were presented for the classical circle polynomials in Sec.~\ref{sec2}. Thus, in Sec.~\ref{sec4}, the Fourier transform of $Z_n^{m,\alpha}$ is computed in terms of Bessel functions and a Weber-Schafheitlin representation of the radial functions $R_n^{m,\alpha}$ is given. In Sec.~\ref{sec5}, the Radon transform of $Z_n^{m,\alpha}$ is computed in terms of the Gegenbauer polynomials $C_n^{\alpha+1}$, and a representation of the radial functions as Fourier coefficients of the periodic function $C_n^{\alpha+1}(\rho\cos\vart)$, $0\leq\vart<2\pi$, for a fixed $\rho\in[0,1)$ is given. Thus a computation scheme of the DCT-type for the radial functions arises. Then, in Sec.~\ref{sec6}, a scaling result is given. This result is somewhat more awkward than in the classical case since the radial functions $R_n^{m,\alpha}(\rho)$ have restrictions on their behaviour as $\rho\uparrow1$ while the scaled radial functions $R_n^{m,\alpha}(\eps\rho)$, to be considered with $0<\eps<1$, do not. Next, in Sec.~\ref{sec7}, the new Zernike functions are expanded in terms of the classical circle polynomials, with an explicit expression for the expansion coefficients. This makes it possible to transfer all forward computation results from the ENZ theory and from the acoustic Nijboer-Zernike (ANZ) theory for the classical case to the new setting in a semi-analytic form. This is useful for those cases that a closed form or a simple semi-analytic form is not available or awkward to find in the new setting. In Secs.~\ref{sec8}--\ref{sec10}, the focus is on generalizing the results from the ANZ theory, as developed in \cite{ref19}--\cite{ref22} to the more general setting. This gives rise in Sec.~\ref{sec8} to a power series representation of the basic integrals that occur when various acoustic quantities are computed from King's integral for the sound pressure (in the case of baffled-piston radiation). An inverse problem, in which the velocity profile is estimated in terms of its expansion coefficients from near-field measurements via Weyl's formula, is considered in Sec.~\ref{sec9}. Finally, in Sec.~\ref{sec10}, the trial functions as used by Streng and Mellow, following Bouwkamp's solution of the diffraction problem for a circular aperture, are compared with the $Z_{2l}^{0,\alpha={\pm}1/2}$. In \cite{ref46}, Mellow and K\"arkk\"ainen consider radiation from a disk with concentric rings, and for this an indefinite integral involving the radial functions $R_{2l}^{0,{\pm}1/2}(\rho)$ is required. This integral is computed in closed form.

\section{Fourier transform of generalized Zernike circle functions} \label{sec4}
\mbox{} \\[-9mm]

In this section, the 2D Fourier transform of $Z_n^{m,\alpha}$ is computed. It is convenient to write here
\beq \label{e33}
Z_n^{m,\alpha}(\nu,\mu)\equiv Z_n^{m,\alpha}(\rho,\vart)~,
\eq
where $\nu+i\mu=\rho\,e^{i\vart}$ with $\rho\geq0$ and $0\leq\vart<2\pi$. \\ \\
{\bf Theorem 4.1.}~~For $\alpha>{-}1$ and integer $n$, $m$ such that $p=\frac12\,(n-|m|)$ is a non-negative integer, there holds
\begin{eqnarray} \label{e34}
& \mbox{} & \il\!\!\il\,e^{2\pi i\nu x+2\pi i\mu y}\,Z_n^{m,\alpha}(\nu,\mu)\,d\nu\,d\mu~= \nonumber \\[3.5mm]
& & =~2\pi\,i^n\,2^{\alpha}(p+1)_{\alpha}\,\frac{J_{n+\alpha+1}(2\pi r)}{(2\pi r)^{\alpha+1}}\,e^{im\varp}~,
\end{eqnarray}
where $x+iy=r\,e^{i\varp}$ with $r\geq0$ and $0\leq\varp<2\pi$. Furthermore,
\beq \label{e35}
\dil_0^1\,R_n^{|m|,\alpha}(\rho)\,J_{|m|}(2\pi\rho r)\,\rho\,d\rho=({-}1)^p\,2^{\alpha}(p+1)_{\alpha}\, \frac{J_{n+\alpha+1}(2\pi r)}{(2\pi r)^{\alpha+1}}~,
\eq
and
\beq \label{e36}
R_n^{|m|,\alpha}(\rho)=({-}1)^p\,2^{\alpha}(p+1)_{\alpha}\,\dil_0^{\infty}\, \frac{J_{n+\alpha+1}(t)\,J_{|m|}(\rho t)}{t^{\alpha}}\,dt~,~~~~~0\leq\rho<1~.
\eq
{\bf Proof.}~~First consider the case that $m\geq0$. It follows from \cite{ref14}, 11.4.33 (Weber-Schafheitlin integral), with
\beq \label{e37}
\mu=n+p+1=m+2p+\alpha+1\,,~~a=1\,,~~\nu=m\,,~~b=\rho\in[0,1)\,,~~\lambda=\alpha
\eq
that
\begin{eqnarray} \label{e38}
& \mbox{} & \il_0^{\infty}\,\frac{J_{n+\alpha+1}(t)\,J_m(\rho t)}{t^{\alpha}}\,dt~= \nonumber \\[3mm]
& & =~\frac{\rho^m\,\Gamma(m+p+1)}{2^{\alpha}\, \Gamma(m+1)\,\Gamma(p+\alpha+1)}\,F({-}p-\alpha,m+p+1\,;\,m+1\,;\,\rho^2)~. \nonumber \\
& & \mbox{}
\end{eqnarray}
Next from \cite{ref14}, 15.3.3 with
\beq \label{e39}
a=m+1+p+\alpha\,,~~~b={-}p\,,~~~c=m+1\,,~~~z=\rho^2
\eq
it follows that
\begin{eqnarray} \label{e40}
& \mbox{} & F({-}p-\alpha,m+p+1\,;\,m+1\,;\,\rho^2)~= \nonumber \\[3mm]
& & =~(1-\rho^2)^{\alpha}\,F({-}p,m+1+\alpha+p\,;\,m+1\,;\,\rho^2)~= \nonumber \\[3mm]
& & =~(1-\rho^2)^{\alpha}\,\frac{p!}{(m+1)_p}\,P_p^{(m,\alpha)}(1-2\rho^2)~,
\end{eqnarray}
where in the last step \cite{ref14}, 15.4.6 with $n=p$, $\alpha=m$, $\beta=\alpha$ and $z=\rho^2$ has been used. Therefore, for general integer $m$, $|m|\leq n$,
\begin{eqnarray} \label{e41}
& \mbox{} & \il_0^{\infty}\,\frac{J_{n+\alpha+1}(t)\,J_{|m|}(\rho t)}{t^{\alpha}}\,dt~= \nonumber \\[3.5mm]
& & =~\frac{({-}1)^p}{2^{\alpha}(p+1)_{\alpha}}\,\rho^{|m|}(1-\rho^2)^{\alpha}\,P_p^{(\alpha,|m|)}(2\rho^2-1) = \frac{({-}1)^p}{2^{\alpha}(p+1)_{\alpha}}\,R_n^{|m|,\alpha}(\rho)~, \nonumber \\
& & \mbox{}
\end{eqnarray}
where $P_k^{(\alpha,\beta)}({-}x)=({-}1)^k\,P_k^{(\beta,\alpha)}({-}x)$ and the definition (\ref{e26}) of $R_n^{|m|,\alpha}(\rho)$ have been used. This establishes (\ref{e36}).

Next,
\begin{eqnarray} \label{e42}
& \mbox{} & \il_0^{\infty}\il_0^{2\pi}\,e^{-2\pi i\nu x-2\pi i\mu y}\,\frac{J_{n+\alpha+1}(2\pi r)} {(2\pi r)^{\alpha+1}}\,e^{im\varp}\,r\,dr\,d\varp~= \nonumber \\[3.5mm]
& & =~2\pi\,i^m\,\il_0^{\infty}\,J_m({-}2\pi r\rho)\,\frac{J_{n+\alpha+1}(2\pi r)}{(2\pi r)^{\alpha+1}}\,r\,dr\,e^{im\vart}~= \nonumber \\[3.5mm]
& & =~\frac{({-}1)^{|m|}}{2\pi}\,\il_0^{\infty}\,\frac{J_{|m|}(t)\,J_{n+\alpha+1}(t)}{t^{\alpha}}\,dt\, e^{im\vart}~,
\end{eqnarray}
where, subsequently, use is made of
\beq \label{e43}
\nu x+\mu y=\rho\,r\cos(\vart-\varp)~,
\eq
\beq \label{e44}
\frac{1}{2\pi}\,\il_0^{2\pi}\,e^{-it\cos\vart}\,e^{im\vart}\,d\vart=i^m\,J_m({-}t)=({-}i)^{|m|}\,J_{|m|}(t)~,
\eq
see \cite{ref14}, Sec.~9.1, and where the substitution $t=2\pi r$ has been used in the last step in (\ref{e42}). Therefore, using (\ref{e41}), the definitions (\ref{e25})--(\ref{e26}) and $n=|m|+2p$, it follows that
\begin{eqnarray} \label{45}
& \mbox{} & \il\!\!\il\,e^{-2\pi i\mu x-2\pi i\mu y}\,\frac{J_{n+\alpha+1}(2\pi r)}{(2\pi r)^{\alpha+1}}\,r\,dr\,d\varp~= \nonumber \\[3mm]
& & =~\frac{({-}i)^n}{2\pi}~\frac{1}{2^{\alpha}(p+1)_{\alpha}}\,Z_n^{m,\alpha}(\rho,\vart)~.
\end{eqnarray}
Then (\ref{e34}) follows by 2D Fourier inversion. Now also (\ref{e35}) follows using the definitions (\ref{e25})--(\ref{e26}) in (\ref{e34}) and proceeding as in (\ref{e42})--(\ref{e44}). \\ \\
{\bf Notes.} \\[-7mm]
\bi{1.0}
\ITEM{1.} The result in (\ref{e34}) generalizes the case $\alpha=0$ in (\ref{e7}) to general $\alpha>{-}1$.
\ITEM{2.} The result in (\ref{e35}) gives the Hankel transform of order $|m|$ of the radial part $R_n^{|m|,\alpha}$, and (\ref{e36}) is what one gets by inverse Hankel transformation of order $|m|$. These two results are reminiscent of, but clearly different from, the results in (\ref{e10}) and (\ref{e11}) of \cite{ref47}.
\ITEM{3.} By the asymptotics of the Bessel functions, see \cite{ref14}, 9.2.1, it is seen that the Fourier transform of $Z_n^{m,\alpha}$ decays as $r^{-\alpha-3/2}$ as $r\pr\infty$.
\ei

\section{Radon transform of generalized Zernike circle functions; integral representation and DCT-formula for radial parts} \label{sec5}
\mbox{} \\[-9mm]

In this section, the Radon transform of the generalized circle functions is expressed in terms of Gegenbauer polynomials. Furthermore, an integral representation involving these Gegenbauer polynomials for the radial parts is proved, and a method of the DCT-type for computation of the radial parts is shown to follow from this integral representation.

The Radon transform of $Z_n^{m,\alpha}$ is given by
\beq \label{e46}
({\cal R}\,Z_n^{m,\alpha})(\tau,\psi)=\il_{l(\tau,\psi)}\,Z_n^{m,\alpha}(\nu,\mu)\,dl
=\il\,R_n^{|m|,\alpha}(\rho(t))\,e^{im\vart(t)}\,dt~,
\eq
with $l(\tau,\psi)$ and $\rho$, $\vart$ given in Fig.~1 for $\tau\geq0$ and $0\leq\psi<2\pi$. Thus
\beq \label{e47}
\rho(t)=(\tau^2+t^2)^{1/2}\,,~~\vart(t)=\psi+\arctan(t/\tau)=\psi+{\rm sgn}(t)\arccos(\tau/\rho(t))~.
\eq
\begin{figure}[htb!]
\begin{center}
\includegraphics[width=0.70\linewidth]{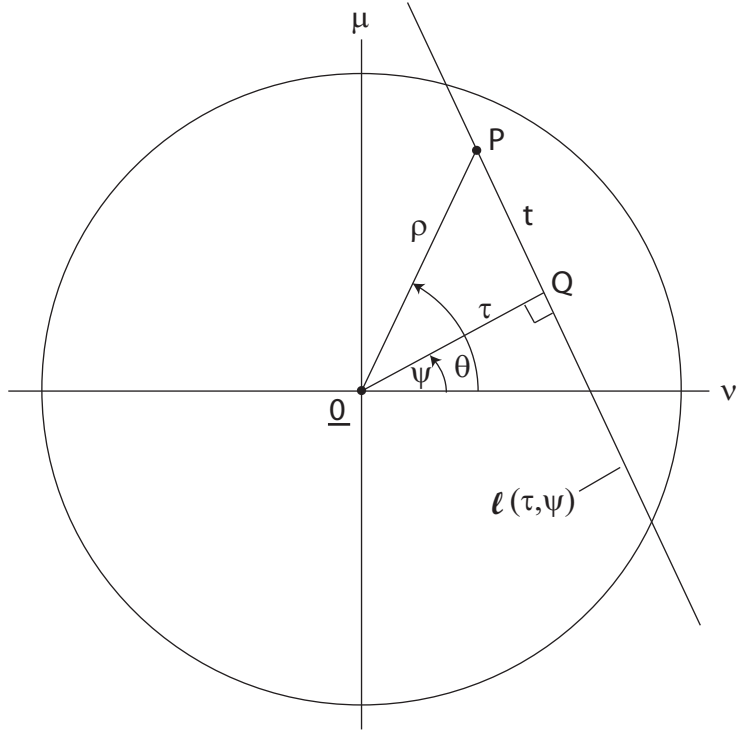}
\caption{Line of integration $l(\tau,\psi)$ and integration point $P=Q+t({-}\sin\psi,\cos\psi)$ with $-\infty<t<\infty$, given in polar coordinates with respect to ${\bf \underline{0}}$ as $(\rho\cos\vart,\rho\sin\vart)$ with $\rho=(\tau^2+t^2)^{1/2}$ and $\vart=\psi+\arctan(t/\tau)$.}  
\end{center}
\end{figure}
Substituting $\rho=\rho(t)$, with $-\infty<t<\infty$ such that $\tau\leq\rho(t)<1$ in the second integral in (\ref{e46}) and using the second form of $\vart(t)$ in (\ref{e47}), it follows that
\begin{eqnarray} \label{e48}
({\cal R}\,Z_n^{m,\alpha})(\tau,\psi) & = & 2e^{im\psi}\,\il_{\tau}^1\,R_n^{|m|,\alpha}(\rho)\cos [m\arccos(\tau/\rho)]\,\frac{\rho\,d\rho}{\sqrt{\rho^2-\tau^2}}~= \nonumber \\[3.5mm]
& = & 2e^{im\psi}\,\il_{\tau}^1\,R_n^{|m|,\alpha}(\rho)\, \frac{T_m(\tau/\rho)}{\sqrt{1-(\tau/\rho)^2}}\,d\rho~,
\end{eqnarray}
where $T_m$ is the Chebyshev polynomial of the first kind and of degree $m$, see \cite{ref14}, Ch.~22. \\ \\
{\bf Theorem 5.1.}~~There holds for $\alpha>{-}1$ and integers $n$, $m$ such that $p=\frac12\,(n-|m|)$ is a non-negative integer that
\beq \label{e49}
({\cal R}\,Z_n^{m,\alpha})(\tau,\psi)= \frac{(p+1)_{\alpha}}{(n+1)_{2\alpha}}~\frac{2^{2\alpha+1}\,\Gamma(\alpha+1)} {n+2\alpha+1}\,(1-\tau^2)^{\alpha+1/2}\,C_n^{\alpha+1}(\tau)\,e^{im\psi}
\eq
when $0\leq\tau\leq1$ and $({\cal R}\,Z_n^{m,\alpha})(\tau,\psi)=0$ for $\tau>1$. Here $C_n^{\alpha+1}$ is the Gegenbauer (or ultraspherical) polynomial corresponding to the weight function $(1-x^2)^{\alpha+1/2}$, $-1<x<1$, and of degree $n$, see \cite{ref14}, Ch.~22. \\[2mm]
{\bf Proof.}~~Consider the case that $m\geq0$ (the case that $m<0$ is obtained from this by complex conjugation). By\cite{ref14}, 11.4.24,
\begin{eqnarray} \label{e50}
\il_{-\infty}^{\infty}\,e^{-i\omega t}\,J_m(t)\,dt & = & \frac{2({-}i)^m\,T_m(\omega)}{(1-\omega^2)^{1/2}}~,~~~~~~ \omega^2<1~, \nonumber \\[2mm]
&= & 0\hspace*{2.5cm},~~~~~~\omega^2>1~,
\end{eqnarray}
and so, from (\ref{e48})
\begin{eqnarray} \label{e51}
({\cal R}\,Z_n^{m,\alpha})(\tau,\psi) & = & e^{im\psi}\,\il_0^1\,R_n^{m,\alpha}(\rho)\,\left( i^m\,\il_{-\infty}^{\infty}\,e^{-i\tau t/\rho}\,J_m(t)\,dt\right)\,d\rho~= \nonumber \\[3.5mm]
& = & e^{im\psi}\,i^m\,\il_{-\infty}^{\infty}\,e^{-i\tau s}\,\left(\il_0^1\,R_n^{m,\alpha}(\rho)\,J_m(\rho s) \,\rho\,d\rho\right)\,ds~= \nonumber \\[3.5mm]
& = & e^{im\psi}\,i^n\,2^{\alpha}(p+1)_{\alpha}\,\il_{-\infty}^{\infty}\,e^{-i\tau s}\, \frac{J_{n+\alpha+1}(s)} {s^{\alpha+1}}\,ds~,
\end{eqnarray}
where (\ref{e35}) has been used, together with $i^m({-}1)^p=i^n$.

By \cite{ref49}, 1.12, item (10), there holds for even $n$ ($a=1$, $2n$ instead of $n$, $\nu=\alpha+1$)
\begin{eqnarray} \label{e52}
& \mbox{} & 2\,\il_0^{\infty}\cos\tau s\,\frac{J_{n+\alpha+1}(s)}{s^{\alpha+1}}\,ds~= \nonumber \\[3.5mm]
& & =~({-}1)^{\frac12 n}\,\frac{2^{\alpha+1}\,n!\,\Gamma(\alpha+1)}{\Gamma(n+2\alpha+2)}\, (1-\tau^2)^{\alpha+1/2} \,C_n^{\alpha+1}(\tau)~,~~~~~~0\leq\tau<1~, \nonumber \\
& \mbox{} &
\end{eqnarray}
while the integral in (\ref{e52}) equals 0 when $\tau>1$. By \cite{ref49}, 2.12, item (10), there holds for odd $n$ ($a=1$, $2n+1$ instead of $n$, $\nu=\alpha+1$)
\begin{eqnarray} \label{e53}
& \mbox{} & -2i\,\il_0^{\infty}\sin\tau s\,\frac{J_{n+\alpha+1}(s)}{s^{\alpha+1}}\,ds~= \nonumber \\[3.5mm]
& & =~{-}i({-}1)^{\frac{n-1}{2}}\,\frac{2^{\alpha+1}\,n!\,\Gamma(\alpha+1)}{\Gamma(n+2\alpha+2)}\, (1-\tau^2)^{\alpha+1/2}\,C_n^{\alpha+1}(\tau)~,~~~~~~0\leq\tau<1~, \nonumber \\
& \mbox{} &
\end{eqnarray}
while the integral in (\ref{e53}) equals 0 when $\tau>1$. Using (\ref{e52}) and (\ref{e53}) in (\ref{e51}), it follows upon some administration with $i^n$ and $({-}1)^{\frac12 n}$, $({-}1)^{\frac12 (n-1)}$ that for $0\leq\tau<1$
\beq \label{e54}
({\cal R}\,Z_n^{m,\alpha})(\tau,\psi)=(p+1)_{\alpha}\, \frac{2^{2\alpha+1}\,n!\,\Gamma(\alpha+1)} {\Gamma(n+2\alpha+2)}\,(1-\tau^2)^{\alpha+1/2}\,C_n^{\alpha+1}(\tau)\,e^{im\psi}~,
\eq
while $({\cal R}\,Z_n^{m,\alpha})(\tau,\psi)=0$ when $\tau>1$. The result follows now upon some further administration with Pochhammer symbols. \\ \\
{\bf Theorem 5.2.}~~There holds for $\alpha>{-}1$ and integer $n$ and $m$ such that $n-|m|$ is even and non-negative
\beq \label{e55}
\Bigl(\!\ba{c} q+\alpha \\ q\ea\!\Bigr)(1-\rho^2)^{-\alpha}\,R_n^{|m|,\alpha}(\rho)=\frac{1}{2\pi}\,\il_0^{2\pi} \,C_n^{\alpha+1}(\rho\cos\vart)\,e^{-im\vart}\,d\vart~,
\eq
where $p=\frac12\,(n-|m|)$, $q=\frac12\,(n+|m|)$. \\[2mm]
{\bf Proof.}~~There holds, taking $|\nu|=\tau$ and $\psi=0$ or $\pi$ according as $\nu \geq 0$ or $\nu < 0$ in (\ref{e49}), with separate consideration of even $m$ and odd $m$ in the case $\nu < 0$,
\beq \label{e56}
\il_{-\sqrt{1-\nu^2}}^{\sqrt{1-\nu^2}}\,Z_n^{m,\alpha}(\nu,\mu)\,d\mu=K_{nm}(1-\nu^2)^{\alpha+1/2}\,C_n^{\alpha+1}(\nu)~,
\eq
where
\beq \label{e57}
K_{nm}=\frac{(p+1)_{\alpha}}{(n+1)_{2\alpha}}~\frac{2^{2\alpha+1}\,\Gamma(\alpha+1)}{n+2\alpha+1}~.
\eq
Expand
\beq \label{e58}
C_n^{\alpha+1}(\nu)=\sum_{n',m'}\,\beta_{nn'}^{mm'}(1-\rho^2)^{-\alpha}\,Z_{n'}^{m',\alpha}(\nu,\mu)
\eq
as a function depending only on $\nu$ with $\nu^2+\mu^2\leq1$. By orthogonality, see (\ref{e31}), there holds
\beq \label{e59}
\beta_{nn'}^{mm'}=L_{n'm'}^{-1}\,\il\hspace*{-6mm}\il_{\nu^2+\mu^2\leq1}\,C_n^{\alpha+1}(\nu)\, Z_{n'}^{m',\alpha} (\nu,\mu)\,d\nu\,d\mu~,
\eq
where
\beq \label{e60}
L_{n'm'}=\frac{\pi}{n'+\alpha+1}~\frac{(p'+1)_{\alpha}}{(p'+|m'|+1)_{\alpha}}~.
\eq
Using (\ref{e56}), it is seen that
\begin{eqnarray} \label{e61}
\beta_{nn'}^{mm'} & = & L_{n'm'}^{-1}\,\il_{-1}^1\,C_n^{\alpha+1}(\nu)\left(\il_{-\sqrt{1-\nu^2}}^{\sqrt{1-\nu^2}}\, Z_{n'}^{m',\alpha}(\nu,\mu)\,d\mu\right)\,d\nu~= \nonumber \\[3.5mm]
& = & L_{n'm'}^{-1}\,K_{n'm'}\,\il_{-1}^1\,(1-\nu^2)^{\alpha+1/2}\,C_n^{\alpha+1}(\nu)\,C_{n'}^{\alpha+1}(\nu)\,d\nu~= \nonumber \\[3.5mm]
& = & L_{n'm'}^{-1}\,K_{n'm'}\,M_n\,\delta_{nn'}~,
\end{eqnarray}
where $M_n$ follows from orthogonality of the $C^{\alpha+1}$ as, see \cite{ref16}, (7.8),
\beq \label{e62}
M_n=\frac{\pi\cdot2^{-2\alpha-1}\cdot\Gamma(n+2\alpha+2)}{n!\,(n+\alpha+1)\,\Gamma^2(\alpha+1)}~.
\eq
With $\nu=\rho\cos\vart$ it then follows that
\begin{eqnarray} \label{e63}
C_n^{\alpha+1}(\rho\cos\vart) & = & \sum_{m'}\,\beta_{nn}^{mm'}(1-\rho^2)^{-\alpha}\,Z_n^{m',\alpha}(\rho,\vart)~= \nonumber \\[3.5mm]
& = & \sum_{m'}\,\beta_{nn}^{mm'}\,\rho^{|m'|}\,P_{\frac{n-|m'|}{2}}^{(\alpha,|m'|)}(2\rho^2-1)\,e^{im\vart}~.
\end{eqnarray}
Therefore, see (\ref{e26}),
\beq \label{e64}
\beta_{nn'}^{mm'}\,\rho^{|m'|}\,P_{\frac{n-|m'|}{2}}^{(\alpha,|m'|)}(2\rho^2-1)=\frac{1}{2\pi}\,\il_0^{2\pi}\, C_n^{\alpha+1}(\rho\cos\vart)\,e^{-im'\vart}\,d\vart
\eq
for integer $m'$ such that $p'':=\frac12\,(n-|m'|)$ is a non-negative integer. An explicit computation from (\ref{e57}), (\ref{e60}), (\ref{e61}) yields now
\begin{eqnarray} \label{e65}
\beta_{nn}^{mm'} & = & L_{nm'}^{-1}\,K_{nm'}\,M_n~= \nonumber \\[3.5mm]
& = & \Bigl(\frac{\pi}{n+\alpha+1}~\frac{(p''+1)_{\alpha}}{(p''+|m'|+1)_{\alpha}}\Bigr)^{-1}\cdot
\frac{(p''+1)_{\alpha}}{(n+1)_{2\alpha}}~\frac{2^{2\alpha+1}\,\Gamma(\alpha+1)}{n+2\alpha+1}~\cdot \nonumber \\[3.5mm]
& & \cdot~
\frac{\pi\cdot2^{-2\alpha-1}\,\Gamma(n+2\alpha+2)}{n!\,(n+\alpha+1)\,\Gamma^2(\alpha+1)}~= \nonumber \\[3.5mm]
& = & \frac{(p''+|m'|+1)_{\alpha}}{\Gamma(\alpha+1)}=\frac{\Gamma(q''+\alpha+1)}{\Gamma(q''+1)\,\Gamma(\alpha+1)} =\Bigl(\!\ba{c} q''+\alpha \\ q \ea\!\Bigr)~,
\end{eqnarray}
in which $q''=p''+|m'|=\frac12\,(n+|m'|)$. Now replace $m'$ by $m$ in (\ref{e64}) and (\ref{e65}), and (\ref{e55}) results. \\ \\
{\bf Notes.} \\[-7mm]
\bi{1.0}
\ITEM{1.} Theorem 5.1 generalizes the case $\alpha=0$ in (\ref{e14}) to general $\alpha>{-}1$. A further generalization, to orthogonal functions on spheres of general dimension $N$ instead of disks, is provided in \cite{refX}, Theorem 3.1. The proof of Theorem~5.1 as given here follows rather closely the approach of Cormack in \cite{ref37}--\cite{ref38} which differs from the approach used in \cite{refX}.
\ITEM{2.} Theorem 5.2 generalizes the case $\alpha=0$ of the integral representation of the $R_n^{|m|}$ in \cite{ref39}, (A.10) to the case of general $\alpha>{-}1$. Furthermore, for a fixed $\rho\in(0,1)$, the formula (\ref{e55}) can be discretized to
\beq \label{e66}
\Bigl(\!\ba{c} q+\alpha \\ q \ea\!\Bigr)(1-\rho^2)^{-\alpha}\,R_n^{|m|,\alpha}(\rho)=
\frac1N\,\sum_{k=0}^N\,C_n^{\alpha+1}\Bigl(\rho\cos\frac{2\pi k}{N}\Bigr)\,e^{-2\pi i\frac{k}{N}}
\eq
when $N$ is an integer $>\:n+|m|$. This yields a method of the DFT-type to compute the radial parts fast and reliably.
\ITEM{3.} The integral representation in (\ref{e55}) is an excellent starting point to derive the asymptotics of the radial parts, by stationary phase methods, etc., when $n,|m|\pr\infty$ such that $n/|m|\pr\kappa\in(0,1)$, with $\alpha$ fixed and $\rho$ bounded away from 0 and 1. See \cite{ref50}, ENZ document, Sec.~7, item~4.
\ei

\section{Scaling theory for generalized Zernike circle functions} \label{sec6}
\mbox{} \\[-9mm]

Scaling theory for generalized Zernike circle functions is compromised by the occurrence of the factor $(1-\rho^2)^{\alpha}$. In the first place, one has to restrict the scaling parameter $\eps$ in $Z_n^{m,\alpha}(\eps\rho,\vart)$ to the range $0\leq\eps<1$. While this restriction is quite natural, the scaling results, such as (\ref{e17}), for the case that $\alpha=0$ allows unrestricted values of $\eps$ due to polynomial form of the radial parts. Furthermore, the value of $Z_n^{m,\alpha}(\eps\rho,\vart)$ at $\rho=1$ (with $0\leq\eps<1$) is in general finite and unequal to 0. This implies that the only natural candidate among the systems $(Z_{n'}^{m',\alpha'} (\rho,\vart))_{n',m'}$ as expansion set for $Z_n^{m,\alpha}(\eps\rho,\vart)$ is the case that $\alpha'=0$. Finally, an extension to shift-and-scaling theory, as in \cite{ref40}, seems also more cumbersome. Nevertheless, for the radial part there is the following result. \\ \\
{\bf Theorem 6.1.}~~Let $\alpha>{-}1$, $0\leq\eps<1$, and let $n$, $m$ be non-negative integers such that $p=\frac12\,(n-m)$ is a non-negative integer. Then
\beq \label{e67}
R_n^{m,\alpha}(\eps\rho)=\sum_{n'=m,m+2,...}\,C_{nn'}^{m,\alpha}(\eps)\,R_{n'}^m(\rho)~,~~~~~~0\leq\rho<1~,
\eq
in which the $C$'s can be expressed as the sum of two hypergeometric functions ${}_2F_1$. Furthermore,
\begin{eqnarray} \label{e68}
& \mbox{} & C_{nn'}^{m,\alpha}(\eps)=\frac{(p+1)_{\alpha}}{(p'+1)_{\alpha}}\,\Bigl(R_n^{n',\alpha}(\eps)-\frac{p'+\alpha}{p'}\, R_n^{n'+2,\alpha}(\eps)\Bigr)~, \nonumber \\[3mm]
& & \hspace*{5.2cm}n'=m,m+2,...,n-2~,
\end{eqnarray}
where $p'=\frac12\,(n-n')$. Finally, when $\alpha$ is a non-negative integer, it holds that
\beq \label{e69}
C_{nn'}^{m,\alpha}(\eps)=0~,~~~~~~n'=n+2\alpha+2,n+2\alpha+4,...~.
\eq
{\bf Proof.}~~By the orthogonality condition (\ref{e31}) for the case $\alpha=0$, there holds for $n'=m,m+2,...$
\beq \label{e70}
C_{nn'}^{m,\alpha}(\eps)=2(n'+1)\,\il_0^1\,R_n^{m,\alpha}(\eps\rho)\,R_{n'}^m(\rho)\,\rho\,d\rho~.
\eq
Now, by (\ref{e36}) and, subsequently (\ref{e35}) with $\alpha=0$, $2\pi r=\eps t$,
\begin{eqnarray} \label{e71}
& \mbox{} & \il_0^1\,R_n^{m,\alpha}(\eps\rho)\,R_{n'}^m(\rho)\,\rho\,d\rho~= \nonumber \\[3.5mm]
& & =~({-}1)^p\,2^{\alpha}(p+1)_{\alpha}\,\il_0^1\,\left(\il_0^{\infty}\, \frac{J_{n+\alpha+1}(t)\,J_m(\eps\rho t)} {t^{\alpha}}\,dt\right)\,R_{n'}^m(\rho)\,\rho\,d\rho~= \nonumber \\[3.5mm]
& & =~({-}1)^p\,2^{\alpha}(p+1)_{\alpha}\,\il_0^{\infty}\,\frac{J_{n+\alpha+1}(t)}{t^{\alpha}}\, \left(\il_0^1\,R_{n'}^m(\rho)\,J_m(\eps\rho t)\,\rho\,d\rho\right)\,dt~= \nonumber \\[3.5mm]
& & =~({-}1)^p\,2^{\alpha}(p+1)_{\alpha}({-}1)^{(n'-m)/2}\,\il_0^{\infty}\,\frac{J_{n+\alpha+1}(t)\, J_{n'+1}(\eps t)} {t^{\alpha}\cdot\eps t}\,dt~.
\end{eqnarray}
Using \cite{ref14}, first item in 9.1.27,
\beq \label{e72}
\frac{J_{n'+1}(z)}{z}=\frac{1}{2(n'+1)}\,(J_{n'}(z)+J_{n'+2}(z))~,
\eq
it is then found that
\beq \label{e73}
C_{nn'}^{m,\alpha}(\eps)=({-}1)^{(n+n'-2m)/2}\,2^{\alpha}(p+1)_{\alpha}(I_{nn'}(\alpha,\eps)+I_{n,n'+2}(\alpha, \eps))~,
\eq
where
\beq \label{e74}
I_{nn''}(\alpha,\eps)=\il_0^{\infty}\,\frac{J_{n+\alpha+1}(t)\,J_{n''}(\eps t)}{t^{\alpha}}\,dt~.
\eq
Now, for $n''=m,m+2,...,n\,$, there holds by (\ref{e35})
\beq \label{e75}
I_{nn''}(\alpha,\eps)=\frac{({-}1)^{p''}}{2^{\alpha}(p''+1)_{\alpha}}\,R_n^{n'',\alpha}(\eps)~,
\eq
where $p''=\frac12\,(n-n'')$. For $n''=n+2,n+4,...\,$, there holds by (\ref{e38})
\beq \label{e76}
I_{nn''}(\alpha,\eps)=\frac{\eps^m\,\Gamma(n''+p''+1)}{2^{\alpha}\,\Gamma(n''+1)\,\Gamma(p''+\alpha+1)} \,F({-}p''-\alpha,n''+p''+1\,;\,n''+1\,;\,\eps^2)~,
\eq
where again $p''=\frac12\,(n-n'')$. When $\alpha$ is a non-negative integer, (\ref{e76}) vanishes when $p''+\alpha$ is a negative integer, i.e., when
\beq \label{e77}
n''=n+2\alpha+2,n+2\alpha+4,...~.
\eq
When $\alpha$ is not an integer, infinitely many of the $I_{nn''}$ must be expected to be non-vanishing.

From (\ref{e73}) and (\ref{e75})it is found for $n'=m,m+2,...,n-2$ that
\begin{eqnarray} \label{e78}
C_{nn'}^{m,\alpha} & = & ({-}1)^{(n+n'-2m)/2}\,2^{\alpha}(p+1)_{\alpha}~\cdot \nonumber \\[3.5mm]
& & \cdot~\left[\frac{({-}1)^{(n-n')/2}}{2^{\alpha}\Bigl(\dfrac{n-n'}{2}+1\Bigr)_{\alpha}}\, R_n^{n',\alpha}(\eps) + \frac{({-}1)^{(n-n'-2)/2}}{2^{\alpha}\Bigl(\dfrac{n-n'}{2}\Bigr)_{\alpha}}\,R_n^{n'+2,\alpha}(\eps)\right]~= \nonumber \\[3.5mm]
& = & \frac{(p+1)_{\alpha}}{(p'+1)_{\alpha}}\,\Bigl(R_n^{n',\alpha}(\eps)-\frac{p'+\alpha}{p'}\, R_n^{n'+2,\alpha} (\eps)\Bigr)~,
\end{eqnarray}
where $p'=\frac12\,(n-n')$ and the proof is complete.

\section{Forward computation schemes for generalized Zernike circle polynomials from ordinary ENZ and ANZ} \label{sec7}
\mbox{} \\[-9mm]

In this section, the generalized Zernike circle functions $Z_n^{m,\alpha}$ are expanded with respect to the system $(Z_{n'}^{m'})_{m',n'}$ of classical circle polynomials. The azimuthal dependence is in all cases through the factor $\exp(im\vart)$, and this allows restriction of the attention to the radial parts only. \\ \\
{\bf Theorem 7.1.}~~Let $\alpha>{-}1$ and let $n$, $m$ be non-negative integers such that $p=\frac12\,(n-m)$ is a non-negative integer. Then
\beq \label{e79}
R_n^{m,\alpha}(\rho)=\sum_{k=0}^{\infty}\,C_k\,R_{m+2k}^m(\rho)~,~~~~~~0\leq\rho<1~,
\eq
where $C_k=0$ for $k=0,1,...,p-1$ and
\begin{eqnarray} \label{e80}
C_k & = & ({-}1)^{k-p}\,\frac{m+2k+1}{m+k+p+\alpha+1}\,\Bigl(\!\ba{c} m+p+k \\ p \ea\!\Bigr) \Bigl(\!\ba{c} \alpha \\ k-p \ea\!\Bigr)\,/ \nonumber \\[3.5mm]
& & /\,\Bigl(\!\ba{c} m+k+p+\alpha \\ m+k \ea\!\Bigr)~= \nonumber \\[3.5mm]
& = & \frac{m+2k+1}{m+k+p+\alpha+1}~\frac{({-}\alpha)_{k-p}}{(k-p)!}~\frac{(p+1)_{\alpha}}{(m+k+p+1)_{\alpha}}
\end{eqnarray}
when $k=p,p+1,...~$. \\[2mm]
{\bf Proof.}~~By the orthogonality condition (\ref{e31}) with $\alpha=0$, there holds
\beq \label{e81}
C_k=2(m+2k+1)\,\il_0^1\,R_n^{m,\alpha}(\rho)\,R_{m+2k}^m(\rho)\,\rho\,d\rho~.
\eq
Using the definition (\ref{e26}) of $R_{n=m+2p}^{m,\alpha}$ and $R_{m+2k}^m$, and using in the integral in (\ref{e81}) the substitution
\beq \label{e82}
x=2\rho^2-1\in[{-}1,1]\,,~~\rho^2=\tfrac12\,(1+x)\,,~~1-\rho^2=\tfrac12\,(1-x)\,,~~\rho\,d\rho=\tfrac14\,dx~,
\eq
this becomes
\begin{eqnarray} \label{e83}
& \mbox{} & \hspace*{-5mm}C_k~= \nonumber \\[3mm]
& & \hspace*{-5mm}=~2(m+2k+1)\,\il_0^1\,\rho^{2m}(1-\rho^2)^{\alpha}\,P_p^{(\alpha,m)}(2\rho^2-1)\,P_k^{(0,m)}(2\rho^2-1)\, \rho\,d\rho~= \nonumber \\[3.5mm]
& & \hspace*{-5mm}=~\frac{m+2k+1}{2^{m+\alpha+1}}\,\il_{-1}^1\,(1-x)^{\alpha}\,(1+x)^m\,P_p^{(\alpha,m)}(x)\,P_k^{(0,m)}(x)\,dx~.
\end{eqnarray}
By Rodriguez' formula, see \cite{ref15}, (4.3.1) or \cite{ref16}, p.~161, there holds
\beq \label{e84}
(1-x)^{\alpha}\,(1+x)^m\,P_p^{(\alpha,m)}(x)=\frac{({-}1)^p}{2^p\,p!}\,\Bigl(\frac{d}{dx}\Bigr)^p\, [(1-x)^{\alpha+p}(1+x)^{m+p}]~.
\eq
Then, by $p$ partial integrations from (\ref{e83}) and (\ref{e84}),
\beq \label{e85}
C_k=\frac{m+2k+1}{2^{m+p+\alpha+1}\,p!}\,\il_{-1}^1\,(1-x)^{p+\alpha}(1+x)^{p+m}\Bigl(\frac{d}{dx}\Bigr)^p\,P_k^{(0,m)} (x)\,dx~.
\eq
Next, by using, see \cite{ref15}, (4.2.17) or \cite{ref16}, (4.14),
\beq \label{e86}
\frac{d}{dx}\,P_n^{(\alpha,\beta)}(x)=\tfrac12\,(n+\alpha+\beta+1)\,P_{n-1}^{(\alpha+1,\beta+1)}(x)
\eq
repeatedly, it is found that for $k\geq p$
\beq \label{e87}
\Bigl(\frac{d}{dx}\Bigr)^p\,P_k^{(0,m)}(x)=\frac{1}{2^p}~\frac{(k+m+p)!}{(k+m)!}\,P_{k-p}^{(p,m+p)}(x)~,
\eq
while this vanishes for $k<p$. Hence, $C_k=0$ for $k<p$ and for $k\geq p$ it is found that
\beq \label{e88}
C_k=\frac{m+2k+1}{2^{m+2p+\alpha+1}}~\frac{(k+m+p)!}{(k+m)!\,p!}\,\il_{-1}^1\,(1-x)^{p+\alpha}(1+x)^{p+m}\, P_{k-p}^{(p,m-p)}(x)\,dx~.
\eq
Next, again by Rodriguez' formula,
\beq \label{e89}
(1-x)^p(1+x)^{m+p}\,P_{k-p}^{(p,m+p)}(x)=\frac{({-}1)^{k-p}}{2^{k-p}(k-p)!}\,\Bigl(\frac{d}{dx}\Bigr)^{k-p} \,[(1-x)^k(1+x)^{k+m}]~.
\eq
Thus, it is found from (\ref{e88}) and (\ref{e89}) by $k-p$ partial integrations that
\begin{eqnarray} \label{e90}
C_k & = & \frac{m+2k+1}{2^{m+k+p+\alpha+1}}~\frac{(k+m+p)!\,({-}1)^{k-p}}{(k+m)!\,p!\,(k-p)!}~\cdot \nonumber \\[3.5mm]
& & \cdot~\il_{-1}^1\,(1-x)^{\alpha}\,\Bigl(\frac{d}{dx}\Bigr)^{k-p}\,[(1-x)^k(1+x)^{k+m}]\,dx~= \nonumber \\[3.5mm]
& = & \frac{m+2k+1}{2^{m+k+p+\alpha+1}}~\frac{(k+m+p)!\,({-}1)^{k-p}}{(k+m)!\,p!\,(k-p)!}\cdot\frac{\Gamma(\alpha+1)} {\Gamma(\alpha-k+p+1)}~\cdot \nonumber \\[3.5mm]
& & \cdot~\il_{-1}^1\,(1-x)^{\alpha+p}(1+x)^{k+m}\,dx~.
\end{eqnarray}
The remaining integral in (\ref{e90}) can be evaluated in terms of $\Gamma$-functions as
\beq \label{e91}
\il_{-1}^1\,(1-x)^{\alpha+p}(1+x)^{k+m}\,dx=2^{m+k+p+\alpha+1}\,\frac{\Gamma(k+m+1)\,\Gamma(p+\alpha+1)} {\Gamma(k+m+p+\alpha+1)}~.
\eq
The final result in (\ref{e80}) then follows upon some further administration with binomials and Pochhammer symbols. \\ \\
{\bf Notes.} \\[-7mm]
\bi{1.0}
\ITEM{1.} When $\alpha$ is a non-negative integer, $C_k$ vanishes for $k\geq p+\alpha$.
\ITEM{2.} For the case that $m=n=0$, so that $p=0$, the expansion of $(1-\rho^2)^{\alpha}$ is obtained, with expansion coefficients
\beq \label{e92}
C_k=({-}1)^k\,\frac{2k+1}{k+1}\,\Bigl(\!\ba{c} \alpha \\ k \ea\!\Bigr)\,/\,\Bigl(\!\ba{c} k+\alpha+1 \\ \alpha \ea\!\Bigr)~,
\eq
compare \cite{ref19}, (10).
\ITEM{3.} A short-cut of the proof of Theorem~7.1 can be obtained by noting that the integral on the last line of (\ref{e83}) is essentially equal to the expansion coefficient $D_p$ in the connection formula
\beq \label{e93}
P_k^{(0,m)}(x)=\sum_{p=0}^k\,D_p\,P_p^{(\alpha,m)}(x)~.
\eq
These connection coefficients are given in \cite{ref51}, Theorem~7.1.3 in terms of Pochhammer symbols. The proof of Theorem~7.1 as given here is ``self-contained'', in the sense that only basic properties of Jacobi polynomials are used, while the proof of \cite{ref51}, Theorem~7.1.3 uses also some more advanced properties of the hypergeometric function ${}_3F_2$.
\ei

The result of Theorem~7.1 gives a means to transfer forward computation schemes from the ordinary ENZ or ANZ theory to the general setting. Below is an example of this. \\ \\
{\bf Theorem 7.2.}~~Let $\alpha>{-}1$, and let $n$, $m$ be non-negative integers such that $n-m$ is even and non-negative. Then the through-focus point-spread function $U_n^{m,\alpha}(r,\varp\,;\,f)$ corresponding to $Z_n^{m,\alpha}(\rho,\vart)$, see (\ref{e6}), is given by
\beq \label{e94}
U_n^{m,\alpha}(r,\varp\,;\,f)=2\pi\,i^m\,e^{im\varp}\,\sum_{k=0}^{\infty}\,C_k\,V_{m+2k}^m(r,f)
\eq
with $V_{m+2k}^m$ given in semi-analytic form in (\ref{e12}) and $C_k$ given in (\ref{e80}). \\[2mm]
{\bf Proof.}~~Just insert the expansion (\ref{e79}) into the integral
\beq \label{e95}
\il_0^1\il_0^{2\pi}\,e^{if\rho^2}\,e^{2\pi i\rho r\cos(\vart-\varp)}\,Z_n^{m,\alpha}(\rho,\vart)\,\rho\,d\rho\,d\vart
\eq
for $U_n^{m,\alpha}$ and use (\ref{e11}). \\ \\
{\bf Notes.} \\[-7mm]
\bi{1.0}
\ITEM{1.} In a similar fashion, the on-axis pressure $p_{2i}^{0,\alpha}((0,0,z),k)$ due to the radially symmetric velocity profile $v(\sigma)=Z_{2i}^{0,\alpha}(\sigma/a,\vart)$, see (\ref{e18}), can be obtained from the on-axis pressures $p_{2j}((0,0,z),k)$ due to $Z_{2j}^0(\sigma/a,0)$, given in (\ref{e20}), and the coefficients $C_j$, case $m=0$, in (\ref{e80}).
\ITEM{2.} The availability of the through-focus point-spread functions $U_n^{m,\alpha}$ per Theorem~7.2 makes it possible to do aberration retrieval, with pupil functions expanded in generalized circle functions, in the same spirit as this is done in the ordinary ENZ theory, see Sec.~\ref{sec2}. Similarly, radially symmetric velocity profiles, expanded into radially symmetric generalized circle functions, can be retrieved with the same approach that is used in the ordinary ANZ theory, see \cite{ref19}, Sec.~V.
\ei

\section{Acoustic quantities for baffled-piston radiation from King's integral with generalized circle functions as velocity profiles} \label{sec8}
\mbox{} \\[-9mm]

In this section various acoustic quantities that arise from baffled-piston radiation with a velocity profile that is expanded into generalized Zernike circle functions are computed in semi-analytic form. The starting point is King's integral, second integral expression in (\ref{e18}), in which $V(u)$ is the Hankel transform (\ref{e19}) of order 0 of $v(\sigma)$. Having expanded $v(\sigma)$ in radially symmetric circle functions $Z_{2j}^{0,\alpha}=R_{2j}^{0,\alpha}$, the Hankel transforms
\begin{eqnarray} \label{96}
V_{2j}^{0,\alpha}(u) & = & \il_0^a\,R_{2j}^{0,\alpha}(\sigma/a)\,J_0(u\sigma)\,\sigma\,d\sigma~= \nonumber \\[3.5mm]
& = & ({-}1)^j\,a^2\,2^{\alpha}(j+1)_{\alpha}\,\frac{J_{2j+\alpha+1}(au)}{(au)^{\alpha+1}}~,~~~~~~j=0,1,...~,
\end{eqnarray}
see (\ref{e35}) arise.

The acoustic quantities considered are \\
-- pressure $p((a,0,0)\,;\,k)$ at edge of the radiator, \\[2mm]
-- reaction force $F=\il_S\,p\,dS$ on the radiator, \\
-- the power output $P=\il_S\,p\,v^{\ast}\,dS$ of the radiator.

By taking $z=0$, $w=a$ in King's integral, it is seen that
\beq \label{e97}
p_{{\rm edge}}=p((a,0,0)\,;\,k)=i\rho_0ck\,\il_0^{\infty}\,\frac{J_0(au)\,V(u)}{(u^2-k^2)^{1/2}}\,u\,du~.
\eq
It was shown in \cite{ref20}, Sec.~III from the integral result
\beq \label{e98}
\il_0^a\,J_0(\sigma u)\,\sigma\,d\sigma=au^{-1}J_1(au)
\eq
by taking $z=0$ in King's integral and integrating over $w$, $0\leq w<a$ that
\beq \label{e99}
F=\il_S\,p\,dS=2\pi i\rho_0cka\,\il_0^{\infty}\,\frac{J_1(au)\,V(u)}{(u^2-k^2)^{1/2}}\,du~.
\eq
It was shown in \cite{ref20}, Sec.~IV, from the representation
\beq \label{e100}
p((\sigma,0,0),k)=i\rho_0ck\,\il_0^{\infty}\, \frac{V(u)}{(u^2-k^2)^{1/2}}\,J_0(\sigma u)\,u\,du~,~~~~~~0\leq\sigma<\infty~,
\eq
as a Hankel transform and by using Parseval's formula for Hankel transforms that
\beq \label{e101}
P=\il_S\,p\,v^{\ast}\,dS=2\pi i\rho_0ck\,\il_0^{\infty}\,\frac{V(u)\,V^{\ast}(u)}{(u^2-k^2)^{1/2}}\,u\,du~.
\eq

By inserting $V=V_{2j}^{0,\alpha}$ into the integrals in (\ref{e97}), (\ref{e99}), (\ref{e101}), it is seen that the integrals
\beq \label{e102}
i\,\il_0^{\infty}\,\frac{J_{m+\beta}(au)\,J_{n+\gamma+1}(au)}{(u^2-k^2)^{1/2}\,u^{\beta+\gamma}}\,du
\eq
arise. Explicitly, in (\ref{e102}) take \\
-- $m=0$, $\beta=0$, $n=2j$, $\gamma=\alpha$ for (\ref{e97}), \\
-- $m=0$, $\beta=1$, $n=2j$, $\gamma=\alpha$ for (\ref{e99}), \\
-- $m=2j_1$, $\beta=\alpha+1$, $n=2j_2$, $\gamma=\alpha$ with $j_1\leq j_2$ for (\ref{e101}). \\
These integrals will now be evaluated as a power series in $ka$ using the method of \cite{ref20}, Appendix~A. \\ \\
{\bf Theorem 8.1.}~~Let $\beta\geq0$, $\gamma>{-}1$ and let $n$, $m$ be non-negative integers such that $n-m$ is non-negative and even. Then
\begin{eqnarray} \label{e103}
& \mbox{} & i\,\il_0^{\infty}\,\frac{J_{m+\beta}(au)\,J_{n+\gamma+1}(au)}{(u^2-k^2)^{1/2}\,u^{\beta+\gamma}}\,du~= \nonumber \\[3.5mm]
& & =~\frac{-({-}1)^p\,a^{2\eps}}{2ka}\,\sum_{l=1}^{\infty}\,\frac{(\tfrac12\,l+\tfrac12)_{\eps}} {(\tfrac12\,l+\eps)_{\delta}}~\frac{({-}\tfrac12\,l+1-\beta)_p({-}\tfrac12\,l+1)_q({-}ika)^l}{\Gamma(\tfrac12\,l+p+\gamma+1) \,\Gamma(\tfrac12\,l+q+2\eps+1)}~, \nonumber \\[1mm]
& \mbox{} &
\end{eqnarray}
where
\beq \label{e104}
\eps=\tfrac12\,(\beta+\gamma)\,,~~\delta=\tfrac12\,(\beta-\gamma)\,,~~p=\tfrac12\,(n-m)\,,~~q=\tfrac12\,(n+m)~.
\eq
{\bf Proof.}~~The plan of the proof is entirely the same as that of the proofs of the results in \cite{ref20}, Appendix~A, and so only the main steps with key intermediate results are given.

With $(u^2-k^2)^{1/2}$ as defined below (\ref{e18}) there holds
\begin{eqnarray} \label{e105}
& \mbox{} & i\,\il_0^{\infty}\,\frac{J_{m+\beta}(au)\,J_{n+\gamma+1}(au)}{(u^2-k^2)^{1/2}\,u^{\beta+\gamma}}\,du~= \nonumber \\[3.5mm]
& & =~\il_0^k\,\frac{J_{m+\beta}(au)\,J_{n+\gamma+1}(au)}{u^{\beta+\gamma}\,\sqrt{k^2-u^2}}\,du+i\,\il_k^{\infty}\, \frac{J_{m+\beta}(au)\,J_{n+\gamma+1}(au)}{u^{\beta+\gamma}\,\sqrt{u^2-k^2}}\,du~= \nonumber \\[3.5mm]
& & =~I_1+I_2~.
\end{eqnarray}
As to the integral $I_1$ in (\ref{e105}), the product of the two Bessel functions is written as an integral, see \cite{ref52}, beginning of \S13.61, and \cite{ref20}, (A8), from $-\infty i$ to $\infty i$, the order of integration is reversed and it is used that
\beq \label{e106}
\il_0^k\,\frac{u^{2q+2s+1}}{\sqrt{k^2-u^2}}\,du=k^{2q+2s+1}\,\frac{\Gamma(q+1+s)\,\Gamma(\tfrac12)} {\Gamma(q+\tfrac32+s)}
\eq
to obtain
\begin{eqnarray} \label{e107}
& \mbox{} & \hspace*{-7mm}I_1=\frac{\tfrac12\,\Gamma(\tfrac12)}{2\pi i}\,(\tfrac12\,a)^{2\eps}\,(\tfrac12\,ka)^{2q}~\cdot \nonumber \\[3.5mm]
& & \hspace*{-7mm}\cdot~\il_{-\infty i}^{\infty i}\,\frac{\Gamma({-}s)\,\Gamma(2q+2\eps+2s+2)\,\Gamma(q+1+s)(\tfrac12\,ka)^{2s+1}} {\Gamma(m+\beta+s+1)\,\Gamma(n+\gamma+s+2)\,\Gamma(2q+2\eps+s+2)\,\Gamma(q+s+\tfrac32)}\,ds~. \nonumber \\
& \mbox{} &
\end{eqnarray}
The choice of the integration contour is such that it has all poles of $\Gamma({-}s)$ on its right and all poles of $\Gamma(2q+2\eps+2s+2)$ on its left (this is possible since $q\geq0$ and $\beta+\gamma>{-}2$). Closing the contour to the right, thereby enclosing all poles of $\Gamma({-}s)$ at $s=j=0,1,...$ with residues $({-}1)^{j+1}/j!$, and using the duplication formula of the $\Gamma$-function to write
\beq \label{e108}
\Gamma(2q+2\eps+2j+2)=\frac{2^{2q+2\eps+2j+1}}{\Gamma(\tfrac12)}\,\Gamma(q+\eps+j+1)\,\Gamma(q+\eps+j+\tfrac32)~,
\eq
it follows that
\begin{eqnarray} \label{e109}
I_1 & = & \tfrac12\,a^{2\eps}\,\sum_{j=0}^{\infty}\,({-}1)^j\,\frac{\Gamma(q+j+1)}{j!}~\frac{\Gamma(q+\eps+j+\tfrac32)} {\Gamma(q+j+\tfrac32)}~\frac{\Gamma(q+\eps+j)}{\Gamma(m+\beta+j+1)}~\cdot \nonumber \\[3.5mm]
& & \cdot~\frac{(ka)^{2(q+j)+1}}{\Gamma(n+\gamma+j+2)\,\Gamma(2q+2\eps+j+2)}~.
\end{eqnarray}
Replacing $j+q+1$ by $j=q+1,q+2,...\,$, it is found after some administration with Pochhammer symbols (such as $(x-q)_q=({-}1)^q(1-x)_q$) that
\beq \label{e110}
I_1=\frac{-({-}1)^p\,a^{2\eps}}{2ka}\,\sum_{j=q+1}^{\infty}\,\frac{(j+\tfrac12)_{\eps}}{(j+\eps)_{\delta}}~\frac{({-}j+1)_q ({-}j+1-\beta)_p({-}ika)^{2j}}{\Gamma(j+p+\gamma+1)\,\Gamma(j+q+2\eps+1)}~.
\eq
Here it may be observed that $({-}j+1)_q=0$ for $j=1,...,q\,$, so that the summation in (\ref{e110}) could start at $j=1$ as well.

As to the integral $I_2$ in (\ref{e105}), again the integral representation for the product of two Bessel functions is used, the integration order is reversed, and it is used that
\beq \label{e111}
\il_k^{\infty}\,\frac{u^{2q+2s+1}}{\sqrt{u^2-k^2}}\,du=\tfrac12\,k^{2q+2s+1}\,\frac{\Gamma(\tfrac12)\, \Gamma({-}q-s-\tfrac12)}{\Gamma({-}q-s)}~.
\eq
This yields
\begin{eqnarray} \label{e112}
& \mbox{} & \hspace*{-7mm}I_2=\tfrac12\,i\,\Gamma(\tfrac12)(\tfrac12\,a)^{2\eps}~\cdot \nonumber \\[3.5mm]
& & \hspace*{-7mm}\cdot\,\il_{-\infty i}^{\infty i}\,\frac{\Gamma({-}s)}{\Gamma({-}q-s)}~\frac{\Gamma(2q+2\eps+2s+2) \,\Gamma({-}q-s-\tfrac12)(\tfrac12\,ka)^{2s+2q+1}} {\Gamma(m+\beta+s+1)\,\Gamma(n+\gamma+s+2)\,\Gamma(2q+2\eps+s+2)}\, ds~. \nonumber \\
& \mbox{} &
\end{eqnarray}
The factor $\Gamma({-}s)/\Gamma({-}q-s)$ is a polynomial since $q=0,1,...\,$, and so the integrand has its poles at $s=j-q-\tfrac12\,$, $j=0,1,...\,$, and at $s={-}r-q-\eps-1$, $r=0,1,...\,$. Since $\eps=(\beta+\gamma)/2>{-}1$, the integration contour can be chosen such that all poles $j-q-\tfrac12\,$, with residues $({-}1)^{j+1}/j!$, lie to the right of it while all poles ${-}r-q-\eps-1$ lie to the left of it. Closing the contour to the right, and using the duplication formula again, to write
\beq \label{e113}
\Gamma(2j+2\eps+1)=\frac{2^{2j+2\eps}}{\Gamma(\tfrac12)}\,\Gamma(j+\eps+\tfrac12)\,\Gamma(j+\eps+1)~,
\eq
it follows that
\begin{eqnarray} \label{e114}
I_2 & = & \tfrac12\,i\,a^{2\eps}\,\sum_{j=0}^{\infty}\,({-}1)^j\,\frac{\Gamma(j+1+\eps)}{\Gamma(j+1)}~\frac{\Gamma({-}j+\tfrac12+q)} {\Gamma({-}j+\tfrac12)}~\frac{\Gamma(j+\eps+\tfrac12)}{\Gamma(j+\beta+\tfrac12-p)}~\cdot \nonumber \\[3.5mm]
& & \cdot~\frac{(ka)^{2j}}{\Gamma(j+\gamma+\tfrac32+p)\,\Gamma(j+2\eps+\tfrac32+q)}~.
\end{eqnarray}
Then some administration with Pochhammer symbols (such as $\Gamma(x)/\Gamma(x-p)=({-}1)^p(1-x)_p$) yields
\beq \label{e115}
I_2=\frac{-({-}1)^p\,a^{2\eps}}{2ka}\,\sum_{j=0}^{\infty}\,\frac{(j+1)_{\eps}} {(j+\eps+\tfrac12)_{\delta}}~\frac{({-}j+\tfrac12)_q({-}j-\beta+\tfrac12)_p({-}ika)^{2j+1}} {\Gamma(j+p+\gamma+\tfrac32)\,\Gamma(j+q+2\eps+\tfrac32)}~.
\eq

The result in (\ref{e103}) is now obtained by adding $I_1$ in (\ref{e110}), with summation starting at $j=1$, and $I_2$ in (\ref{e115}), while observing that the terms $j$ in (\ref{e110}) yield the terms in (\ref{e103}) with even $l=2j$, $j=1,2,...\,$, and that the terms $j$ in (\ref{e115}) yield the terms in (\ref{e103}) with odd $l=2j+1$, $j=0,1,...\,$.

\section{Estimating generalized velocity profiles in baffled-piston radiation from near-field pressure data via Weyl's formula} \label{sec9}
\mbox{} \\[-9mm]

In this section a brief sketch is given of how one can estimate, in the setting of baffled-piston radiation, a not necessarily radially symmetric velocity profile from near-field pressure data. The starting point is the Rayleigh integral for the pressure, first integral expression in (\ref{e18}), that is written in normalized form as
\beq \label{e116}
p(\nu,\mu\,;\,\zeta)=\il_{~S}\hspace*{-2mm}\il\,v(\nu',\mu')\,\frac{e^{ikar'}}{r'}\,d\nu'\,d\mu'~,
\eq
where
\beq \label{e117}
r'=((\nu-\nu')^2+(\mu-\mu')^2+\zeta^2)^{1/2}
\eq
is the distance from the field point $(\nu,\mu,\zeta)$, with $\zeta\geq0$, to the point $(\nu',\mu',0)$ on the radiating surface $S$, for which we take the unit disk. For a fixed value of $\zeta>0$, the equation (\ref{e116}) can be written as
\beq \label{e118}
p(\nu,\mu\,;\,\zeta)=(v\,{\ast\ast}\,W({\cdot},{\cdot}\,;\,\zeta))(\nu,\mu)~,
\eq
where $\ast\ast$ denotes $2D$ convolution and
\beq \label{e119}
W(\nu,\mu\,;\,\zeta)=\frac{\exp[ika(\zeta^2+\nu^2+\mu^2)^{1/2}]}{(\zeta^2+\nu^2+\mu^2)^{1/2}}~.
\eq
Using the Fourier transform ${\cal F}$, defined as
\beq \label{e120}
({\cal F}q)(x,y)=\il_{-\infty}^{\infty}\il_{-\infty}^{\infty}\,e^{2\pi i\nu x+2\pi i\mu y}\,q(\nu,\mu)\,d\nu\,d\mu~,~~~~~~x,y\in\dR~,
\eq
the formula (\ref{e118}) can be written as
\beq \label{e121}
{\cal F}\,[p({\cdot},{\cdot}\,;\,\zeta)]={\cal F}v\cdot{\cal F}\,[W({\cdot},{\cdot}\,;\,\zeta)]~.
\eq
By Weyl's result on the representation of spherical waves, see \cite{ref18}, Sec.~13.2.1, there holds
\beq \label{e122}
{\cal F}\,[W({\cdot},{\cdot}\,;\,\zeta)](x,y)=2\pi i\, \frac{\exp\,[i\zeta((ka)^2-(2\pi x)^2-(2\pi y)^2)^{1/2}]} {((ka)^2-(2\pi x)^2-(2\pi y)^2)^{1/2}}~,
\eq
with the same definition of the square root as the one that was used in connection with King's integral in (\ref{e18}).

Now assume that the unknown velocity profile vanishes outside the unit disk and that it has a $(1-\nu^2-\mu^2)^{\alpha}$-behaviour at the edge of the unit disk, where $\alpha>{-}1$. Then $v$ has an expansion
\beq \label{e123}
v(\nu,\mu)=\sum_{n,m}\,C_n^{m,\alpha}\,Z_n^{m,\alpha}(\nu,\mu)
\eq
in generalized Zernike circle functions, with Fourier transform
\beq \label{e124}
{\cal F}v=\sum_{n,m}\,C_n^{m,\alpha}\,{\cal F}\,Z_n^{m,\alpha}~,
\eq
in which ${\cal F}\,Z_n^{m,\alpha}$ is given explicitly in Sec.~\ref{sec4}. Thus, when the pressure $p$ is measured in the near-field plane $(\nu,\mu\,;\,\zeta)$ with $\zeta$ fixed, one can estimate $v$ on the level of its expansion coefficients $C_n^{m,\alpha}$ by adopting a matching approach in (\ref{e121}), using Weyl's result in (\ref{e122}) and the result of Sec.~\ref{sec4} in (\ref{e124}). \\ \\

\section{Comparison with trial functions as used in acoustic design by Mellow and K\"arkk\"ainen} \label{sec10}
\mbox{} \\[-9mm]

Mellow and K\"arkk\"ainen are concerned with design problems in acoustic radiation from a resilient disk (radius $a$) in an infinite or finite baffle ($z=0$), see \cite{ref45}--\cite{ref46}. The front and rear pressure distributions $p_+$ and $p_-$, $p_-={-}p_+$, are assumed to be radially symmetric and to have the form
\beq \label{e125}
\sum_{l=0}^{\infty}\,a_l(1-(\sigma/a)^2)^{l+1/2}
\eq
on the disk in accordance with the choice of trial functions used by Streng \cite{ref42} which is based on the work of Bouwkamp \cite{ref43}. In the case that the normal gradient of the pressure at $z=0$ is considered, as is done by Mellow in \cite{ref44}, an expansion of the form
\beq \label{e126}
\sum_{l=0}^{\infty}\,b_l(1-(\sigma/a)^2)^{l-1/2}
\eq
on the disk has to be considered.

In the design problem considered in \cite{ref45}--\cite{ref46}, the coefficients $a_l$ in (\ref{e125}) are to be found such that the pressure gradient $\frac{\partial p}{\partial z}\,(w,z=0{+})$ equals a desired function $\Phi(w)$ of the distance $w$ of a point in the baffle plane to the origin. In the design problem considered in \cite{ref44}, the coefficients $b_l$ in (\ref{e126}) are to be found such that $p(w,z=0{+})$ equals a desired function $\Psi(w)$.

The pressure $p(w,z)$, $z\geq0$ can be expressed in terms of the boundary data $p_+$, $p_-$ via the dipole version of King's integral. Similarly, via the common version of King's integral, the pressure $p(w,z)$, $z\geq0$, can be expressed in terms of the normal gradient $\frac{\partial p}{\partial z}\,(w,0)$ of $p$ at $z=0$. Inserting the series expansion (\ref{e125}) and (\ref{e126}) into the appropriate version of King's integral, the integrals
\beq \label{e127}
\il_0^{\infty}\,\Bigl(\frac{1}{\mu}\Bigr)^{l\pm 1/2}\,J_0(w\mu)\,J_{l\pm 1/2+1}(a\mu)\,\sigma^{\pm 1}\,d\mu
\eq
arise where $\sigma={-}i(\mu^2-k^2)^{1/2}$ and where the $\pm$ follows the sign choice in the exponent $l\pm 1/2$ of $(1-(\sigma/a)^2)$ in (\ref{e125}) and (\ref{e126}). 
To obtain (\ref{e127}), an explicit result, due to Sonine, for the Hankel transform of order 0 of the functions $(1-(\sigma/a)^2)^{l\pm 1/2}$ has been used. The integrals in (\ref{e127}) are evaluated in the form of a double power series in $ka$ and $w/a$ in \cite{ref44}--\cite{ref46}. Thus, having the pressure available in this semi-analytic form, comprising the coefficients $a_l$ or $b_l$, one can evaluate $\frac{\partial p}{\partial z}\,(w,z=0{+})$ and $p(w,z=0{+})$ and find the coefficients by requiring a best match
 with the desired function $\Phi(w)$ or $\Psi(w)$.

 In the approach of the present paper, the starting point would be an expansion of the form
 \beq \label{e128}
 \sum_{l=0}^{\infty}\,c_l\,R_{2l}^{0,{\pm} 1/2}(\sigma/a)
 \eq
 of the pressure ($+$-sign) or pressure gradient ($-$-sign) on the disk. Following the approach in \cite{ref44}--\cite{ref46}, using either form of King's integral, this gives rise to the integrals
 \beq \label{e129}
 \il_0^{\infty}\,\Bigl(\frac{1}{\mu}\Bigr)^{\pm 1/2}\,J_0(w\mu)\,J_{2l\pm1/2+1}(a\mu)\,\sigma^{\pm 1}\,d\mu~,
 \eq
 where now the result of Sec.~\ref{sec4} on the Hankel transform of $R_{2l}^{0,{\pm}1/2}$ has been used. The integral in (\ref{e129}) is of the same type as the one in (\ref{e127}) and can be evaluated by the method given in \cite{ref44}--\cite{ref46}.

\subsection{Numerical considerations} \label{subsec10.1}
\mbox{} \\[-9mm]

In either approach, it is required to find coefficients such that a best match occurs between the semi-analytically computed pressure gradient or pressure at $z=0{+}$, comprising the coefficients, and the desired functions $\Phi$ or $\Psi$. For any $L=1,2,...\,$, the linear span of the function systems
\beq \label{e130}
\{(1-(\sigma/a)^2)^{l\pm 1/2}\,|\,l=0,1,...,L-1\}
\eq
and
\beq \label{e131}
\{R_{2l}^{0,{\pm}1/2}(\sigma/a)\,|\,l=0,1,...,L-1\}
\eq
is the same. So matching using the first $L$ functions in (\ref{e125}), (\ref{e126}) yields the same result for the best matching pressure gradient or pressure at $z=0{+}$ as matching using the first $L$ functions in (\ref{e128}), in theory. For small values of $L$, one finds numerically practically the same result when either system in (\ref{e130}), (\ref{e131}) is used. In the case that large values $L$ of the number of coefficients to be matched are required, the approach based on (\ref{e125}), (\ref{e126}) is expected to experience numerical problems while the one based on (\ref{e129}) is likely not to have such problems. This is due to the fact that the functions in (\ref{e130}) are nearly linearly dependent while the ones in (\ref{e131}), due to orthogonality, are not, and this is expected to remain so after the linear transformation associated with either version of King's integral. Furthermore, it is to be expected that the semi-analytic forms, used in the matching procedure, that arise from any of the terms $(1-(\sigma/a)^2)^{l\pm1/2}$ must be used with much higher truncation levels than those that arise from the terms $R_{2l}^{0,{\pm}1/2}(\sigma/a)$.

All this can be illustrated by comparing the expansion coefficients of a $(1-(\sigma/a)^2)^{k\pm1/2}$ with respect to the system in (\ref{e131}) with those of an $R_{2p}^{0,{\pm}1/2}(\sigma/a)$ with respect to the system in (\ref{e130}). There is the following general result. \\ \\
{\bf Theorem 10.1.}~~For $m=0,1,...\,$, $\alpha>{-}1$ and $k,p=0,1,...$ there holds
\beq \label{e132}
\rho^m(1-\rho^2)^{k+\alpha}\,e^{im\vart}=\sum_{l=0}^k\,D_{m+2l,k}^{m,\alpha}\,Z_{m+2l}^{m,\alpha}(\rho,\vart)~,
\eq
\beq \label{e133}
Z_{m+2p}^{m,\alpha}(\rho,\vart)=\sum_{r=0}^p\,E_{r,m+2p}^{m,\alpha}\,\rho^m(1-\rho^2)^{r+\alpha}\,e^{im\alpha}~,
\eq
where
\begin{eqnarray} \label{e134}
& \mbox{} & D_{m+2l,k}^{m,\alpha}=\frac{(\alpha+1)_k}{(m+\alpha+1)_k}~\frac{m+2l+\alpha+1}{m+k+l+\alpha+1}~\frac{({-}k)_l(m+\alpha+1)_l} {(\alpha+1)_l(m+k+\alpha+1)_l}~, \nonumber \\[2mm]
& \mbox{} &
\end{eqnarray}
\beq \label{e135}
E_{r,m+2p}^{m,\alpha}=\frac{(\alpha+1)_p}{(1)_p}~\frac{({-}p)_r(m+p+\alpha+1)_r}{(\alpha+1)_r(1)_r}~.
\eq
\mbox{} \\
{\bf Proof.}~~The proof of (\ref{e132}), (\ref{e134}) is quite similar to the one of Theorem~7.1, and so only the main steps and key intermediate results are given. By orthogonality, see (\ref{e31}), and the substitution in (\ref{e82}), there holds
\beq \label{e136}
D_{m+2l,k}^{m,\alpha}=\frac{m+2l+\alpha+1}{2^{m+k+\alpha+1}}~\frac{(l+m+1)_{\alpha}}{(l+1)_{\alpha}}\,\il_{-1}^1\, (1-x)^{k+\alpha}(1+x)^m\,P_l^{(\alpha,m)}(x)\,dx~.
\eq
Next, Rodriguez' formula, (see (\ref{e84})), is used, and $l$ partial integrations are performed. There results for $k\geq l$
\begin{eqnarray} \label{e137}
& \mbox{} & \frac{1}{2^{m+k+\alpha+1}}\,\il_{-1}^1\,(1-x)^{k+\alpha}(1+x)^m\,P_l^{(\alpha,m)}(x)\,dx~= \nonumber \\[3.5mm]
& & =~\frac{({-}1)^l}{2^{m+k+l+\alpha+1}}~\frac{\Gamma(k+l)}{l!\,\Gamma(k+1-l)}\,\il_{-1}^1\, (1-x)^{k+\beta}(1+x)^{l+m}\,dx~,
\end{eqnarray}
while this vanishes for $k<l$. The remaining integral can be expressed in terms of $\Gamma$-functions as in (\ref{e91}), and then the result follows upon some administration with Pochhammer symbols.

From the definition of $Z_{m+1p}^{m,\alpha}$, one should find the $E$'s according to
\beq \label{e138}
P_p^{(\alpha,m)}(2\rho^2-1)=\sum_r\,E_{r,m+2p}^{m,\beta}(1-\rho^2)^r~.
\eq
By \cite{ref14}, 15.4.6, it holds that
\begin{eqnarray} \label{e139}
P_p^{(\alpha,m)}(2\rho^2-1) & = & P_p^{(\alpha,m)}(1-2(1-\rho^2))~= \nonumber \\[3mm]
& = & \frac{(\alpha+1)_p}{p!}\,F({-}p,\alpha+1+m+p\,;\,\alpha+1\,;\,1-\rho^2)~, \nonumber \\
& \mbox{} &
\end{eqnarray}
and then the $E$'s follow from the definition of $F$ in \cite{ref14}, 15.1.1. \\ \\
{\bf Example.}~~For the case $m=0$, it is found that the ratio of the $r^{{\rm th}}$ coefficient for $(1-\rho^2)^{p+\alpha}$ and the $r^{{\rm th}}$ coefficient for $Z_{m+2p}^{0,\alpha}(\rho,\vart)$ satisfy
\beq \label{e140}
\frac{D_{2r,p}^{0,\alpha}}{E_{r,2p}^{0,\alpha}}=\frac{(1)_p}{(\alpha+1)_p}~\frac{2r+\alpha+1}{p+r+\alpha+1}~ \frac{(1)_r(\alpha+1)_r}{((p+\alpha+1)_r)^2}~,
\eq
showing that the $E$'s are much larger than the $D$'s. Note also that $(1-\rho^2)^{p+\alpha}$ and $Z_{m+2p}^{0,\alpha}(\rho,\vart)$ have the same $L^2$-norm, where the weight function $(1-\rho^2)^{-\alpha}$ on the unit disk is used.

\subsection{Indefinite integrals for multi-ring design} \label{subsec10.2}
\mbox{} \\[-9mm]

In \cite{ref46}, Mellow and K\"arkk\"ainen replace the disk by a ring, and in \cite{ref46}, Subsec.~II.F, the total radiation force is expressed as an integral over this ring of $p_+-p_-$. This integral can be expressed explicitly in terms of the $a_l$ in (\ref{e125}) since the functions $(1-(\sigma/a)^2)^{l+1/2}$ have an analytic result for their integrals over a concentric ring. In the case that expansions involving $R_{2l}^{0,{\pm}1/2}$ are used, non-trivial integrals arise. \\ \\
{\bf Theorem 10.2.}~~There holds for $l=0,1,...$
\begin{eqnarray} \label{e142e143}
\il\,R_{2l}^{0,{-}1/2}(\rho)\,\rho\,d\rho & = & \frac{-1}{4l+1}\,(R_{2l}^{0,1/2}(\rho)+R_{2l-2}^{0,1/2}(\rho))~, \\[3.5mm]
\il\,R_{2l}^{0,1/2}(\rho)\,\rho\,d\rho & = & \frac{1}{4l+1}\,\Bigl[\frac{2l+2}{4l+5}\,R_{2l+2}^{0,1/2}(\rho)~+\Bigr. \nonumber \\[3.5mm]
& & -~\Bigl.\frac{4l+3}{(4l+5)(4l+1)}\,R_{2l}^{0,1/2}(\rho)-\frac{2l+1}{4l+1}\,R_{2l-2}^{0,1/2}(\rho)\Bigr]~, \nonumber \\
& \mbox{} &
\end{eqnarray}
where $R_{-2}^{0,1/2}(\rho)\equiv0$ has been set for the case $l=0$. \\[2mm]
{\bf Proof.}~~There holds by \cite{ref15}, $\alpha=0$ in the first item in (4.1.5),
\beq \label{e144}
P_{2l}(x)=({-}1)^l P_l^{({-}1/2,0)}(1-2x^2)~,
\eq
where $P_{2l}$ denotes the Legendre polynomial of degree $2l$. Using this with $x=(1-\rho^2)^{1/2}$, it is seen that
\begin{eqnarray} \label{e145}
R_{2l}^{0,{-}1/2}(\rho) & = & (1-\rho^2)^{-1/2}\,P_l^{({-}1/2,0)}(2\rho^2-1)~= \nonumber \\[3mm]
& = & ({-}1)^l(1-\rho^2)^{-1/2}\,P_{2l}((1-\rho^2)^{1/2})~.
\end{eqnarray}
For $l=0$, it then follows that
\beq \label{e146}
\il\,R_0^{0,{-}1/2}(\rho)\,\rho\,d\rho=\il\,(1-\rho^2)^{-1/2}\,\rho\,d\rho={-}(1-\rho^2)^{1/2}={-}R_0^{0,1/2}(\rho)~.
\eq
For $l=1,2,...$ it follows from (\ref{e145}) and the substitution $\tau=(1-\rho^2)^{1/2}$, $\rho\,d\rho={-}\tau\,d\tau$ that
\beq \label{e147}
\il\,R_{2l}^{0,{-}1/2}(\rho)\,\rho\,d\rho=({-}1)^{l+1}\,\il\,P_{2l}(\tau)\,d\tau~.
\eq
Then from \cite{ref16}, (10.10),
\beq \label{e148}
P_k(\tau)=\frac{1}{2k+1}\,(P_{k+1}'(\tau)-P_{k-1}'(\tau))~,
\eq
it follows that
\beq \label{e149}
\il\,R_{2l}^{0,{-}1/2}(\rho)\,\rho\,d\rho=\frac{({-}1)^{l+1}}{4l+1}\,(P_{2l+1}(\tau)-P_{2l-1}(\tau))~.
\eq
Next, by \cite{ref15}, second item in (4.1.5),
\beq \label{e150}
P_{2l+1}(x)=({-}1)^l\,x\,P_l^{(1/2,0)}(1-2x^2)~.
\eq
Using this with $x=(1-\rho^2)^{1/2}$, it is seen that
\begin{eqnarray} \label{e151}
R_{2l}^{0,1/2}(\rho) & = & (1-\rho^2)^{1/2}\,P_l^{(1/2,0)}(2\rho^2-1)~= \nonumber \\[3mm]
& = & ({-}1)^l\,P_{2l+1}((1-\rho^2)^{1/2})~.
\end{eqnarray}
Now (141) follows from (\ref{e150}) by using $\tau=(1-\rho^2)^{1/2}$.

To show (142), let $l=1,2,...\,$. It follows from (\ref{e151}) and the substitution $\tau=(1-\rho^2)^{1/2}$, $\rho\,d\rho={-}\tau\,d\tau$ that
\beq \label{e152}
\il\,R_{2l}^{0,1/2}(\rho)\,\rho\,d\rho=({-}1)^{l+1}\,\il\,\tau\,P_{2l+1}(\tau)\,d\tau~.
\eq
Then by (\ref{e148}) and partial integration
\beq \label{e153}
\il\,R_{2l}^{0,1/2}(\rho)\,\rho\,d\rho=\frac{({-}1)^{l+1}}{4l+3}\, \Bigl[\tau\,P_{2l+2}(\tau)-\tau\, P_{2l}(\tau)-\il\,(P_{2l+2}(\tau)-P_{2l}(\tau))\,d\tau\Bigr]~.
\eq
Next, \cite{ref16}, (10.2),
\beq \label{e154}
x\,P_k(x)=\frac{k}{2k+1}\,P_{k-1}(x)+\frac{k+1}{2k+1}\,P_{k+1}(x)~,
\eq
is used. It follows from (\ref{e153}) and (\ref{e148}), (\ref{e154}) and some administration that
\begin{eqnarray} \label{e155}
& \mbox{} & \hspace*{-4mm}\il\,R_{2l}^{0,1/2}(\rho)\,\rho\,d\rho~= \nonumber \\[3.5mm]
& & \hspace*{-4mm}=~\frac{({-}1)^{l+1}}{4l+3}\,
\Bigl[\frac{2l+2}{4l+5}\,P_{2l+3}(\tau)+\frac{4l+3}{(4l+5)(4l+1)}\, P_{2l+1}(\tau)-\frac{2l+1}{4l+1}\,P_{2l-1} (\tau)\Bigr]~. \nonumber \\
& \mbox{} &
\end{eqnarray}
Then (142) follows for $l=1,2,...$ by using (\ref{e151}) and $\tau=(1-\rho^2)^{1/2}$. Finally, for $l=0$ it follows from (\ref{e151}) that
\beq \label{e156}
\il\,R_0^{0,1/2}(\rho)\,\rho\,d\rho=\il\,(1-\rho^2)^{1/2}\,\rho\,d\rho={-}\tfrac13\,(1-\rho^2)^{3/2}~,
\eq
while (142) with $l=0$ equals by (\ref{e155})
\beq \label{e157}
-\tfrac13\,[\tfrac35\,P_1(\tau)+\tfrac25\,P_3(\tau)]={-}\tfrac13\,\tau^3={-}\tfrac13\,(1-\rho^2)^{3/2}~.
\eq
The proof is complete. \\ \\
{\bf Acknowledgement.}~~The author wishes to thank R.\ Aarts, J.\ Braat, S.\ van Haver and T.\ Mellow for stimulating discussions and comments.

\end{document}